\documentclass[%
 reprint,superscriptaddress,
 amsmath,amssymb,prx]{revtex4-2}
\usepackage{physics}
\usepackage{xcolor}
\usepackage{xtab,afterpage,longtable}
\usepackage{lipsum}
\usepackage{graphicx}
\usepackage{dcolumn}
\usepackage{booktabs} 
\usepackage{multirow}
\usepackage{color}
\usepackage{hyperref}

\begin{document}

\title{Multi-task learning for molecular electronic structure approaching coupled-cluster accuracy}

\author{Hao Tang}%
 \affiliation{Department of Materials Science and Engineering, Massachusetts Institute of Technology, MA 02139, USA}

\author{Brian Xiao}%
 \affiliation{Department of Physics, Massachusetts Institute of Technology, MA 02139, USA}

\author{Wenhao He}%
   \affiliation{
   The Center for Computational Science and Engineering, Massachusetts Institute of Technology, Cambridge, MA 02139, USA}

\author{Pero Subasic}%
\affiliation{
   Honda Research Institute USA, Inc., 
   San Jose, CA 95134, USA}

\author{Avetik R. Harutyunyan}%
\affiliation{
   Honda Research Institute USA, Inc., 
   San Jose, CA 95134, USA}

\author{Yao Wang}%
\affiliation{
   Department of Chemistry, Emory University, Atlanta, GA 30322, USA}

\author{Fang Liu}%
\affiliation{
   Department of Chemistry, Emory University, Atlanta, GA 30322, USA}

\author{Haowei Xu}%
\email{haoweixu@mit.edu}
\affiliation{
   Department of Nuclear Science and Engineering, Massachusetts Institute of Technology, Cambridge, MA 02139, USA}

\author{Ju Li}%
 \email{liju@mit.edu}
 \affiliation{Department of Materials Science and Engineering, Massachusetts Institute of Technology, MA 02139, USA}
\affiliation{
   Department of Nuclear Science and Engineering, Massachusetts Institute of Technology, Cambridge, MA 02139, USA}

\date{\today}
             
\begin{abstract}

Machine learning (ML) plays an important role in quantum chemistry, providing fast-to-evaluate predictive models for various properties of molecules. However, most existing ML models for molecular electronic properties use density functional theory (DFT) databases as ground truth in training, and their prediction accuracy cannot surpass that of DFT. In this work, we developed a unified ML method for electronic structures of organic molecules using the gold-standard CCSD(T) calculations as training data. Tested on hydrocarbon molecules, our model outperforms DFT with the widely-used hybrid and double hybrid functionals in computational costs and prediction accuracy of various quantum chemical properties. As case studies, we apply the model to aromatic compounds and semiconducting polymers on both ground state and excited state properties, demonstrating its accuracy and generalization capability to complex systems that are hard to calculate using CCSD(T)-level methods. 

\end{abstract}

\maketitle

\section{Introduction}
Computational methods for molecular and condensed matter systems play essential roles in physics, chemistry, and materials science, which can reveal underlying mechanisms of diverse physical phenomena and accelerate materials design~\cite{Yip2005,carter2008challenges}. Among various types of computational methods, quantum chemistry calculations of electronic structure are usually the bottleneck, limiting the computational speed and scalability~\cite{kulik2022roadmap}. In recent years, machine learning (ML) methods have been successfully applied to accelerate molecular dynamics simulations and to improve their accuracy in many application scenarios~\cite{dral2022quantum,TakamotoIL22,TakamotoOLL23}. Particularly, ML inter-atomic potential can predict energy and force of molecular systems with significantly lower computational costs compared to quantum chemistry methods~\cite{TakamotoIL22,TakamotoOLL23,batzner20223, zhang2018deep}. Indeed, recent advances in universal ML potential enable large-scale molecular dynamics simulation with the complexity of realistic physical systems~\cite{merchant2023scaling,chen2022universal,takamoto2022towards,smith2019approaching,zheng2021artificial}. Besides ML inter-atomic potential,  rapid advances also appear in another promising direction, namely the ML density functional, which focuses on further improving the energy prediction towards the chemical accuracy (1 kcal/mol)~\cite{kirkpatrick2021pushing,pederson2022machine,bogojeski2020quantum, bystrom2022cider}.

Besides energy and force, other electronic properties that explicitly involve the electron degrees of freedom are also essential in molecular simulations~\cite{helgaker2013molecular}. In the past few years, ML methods have also been extended to electronic structure of molecules, predicting various electronic properties such as electric multipole moments~\cite{schutt2019unifying,shao2023machine,feng2023accurate}, electron population~\cite{fan2022obtaining}, excited states properties~\cite{cignoni2023electronic,dral2021molecular}, as well as the electronic band structure of condensed matter~\cite{li2022deep,gong2023general}. Most of these methods take the density functional theory (DFT) results as the training data, using neural networks (NN) to fit the single-configurational representation (either the Kohn-Sham Hamiltonian or molecular orbitals) of the DFT calculations~\cite{schutt2019unifying, li2022deep,cignoni2023electronic,unke2021se}. Along with the rapid advances of ML techniques, the NN predictions match the DFT results increasingly well, approaching the chemical accuracy~\cite{shao2023machine,merchant2023scaling}. However, as a mean-field level theory, DFT calculations themselves induce a systematic error, which is usually several times larger than the chemical accuracy~\cite{mardirossian2017thirty}, limiting the overall accuracy of the ML model trained on DFT datasets. 

\begin{figure*}[t]
\centering
\includegraphics[width=\linewidth]{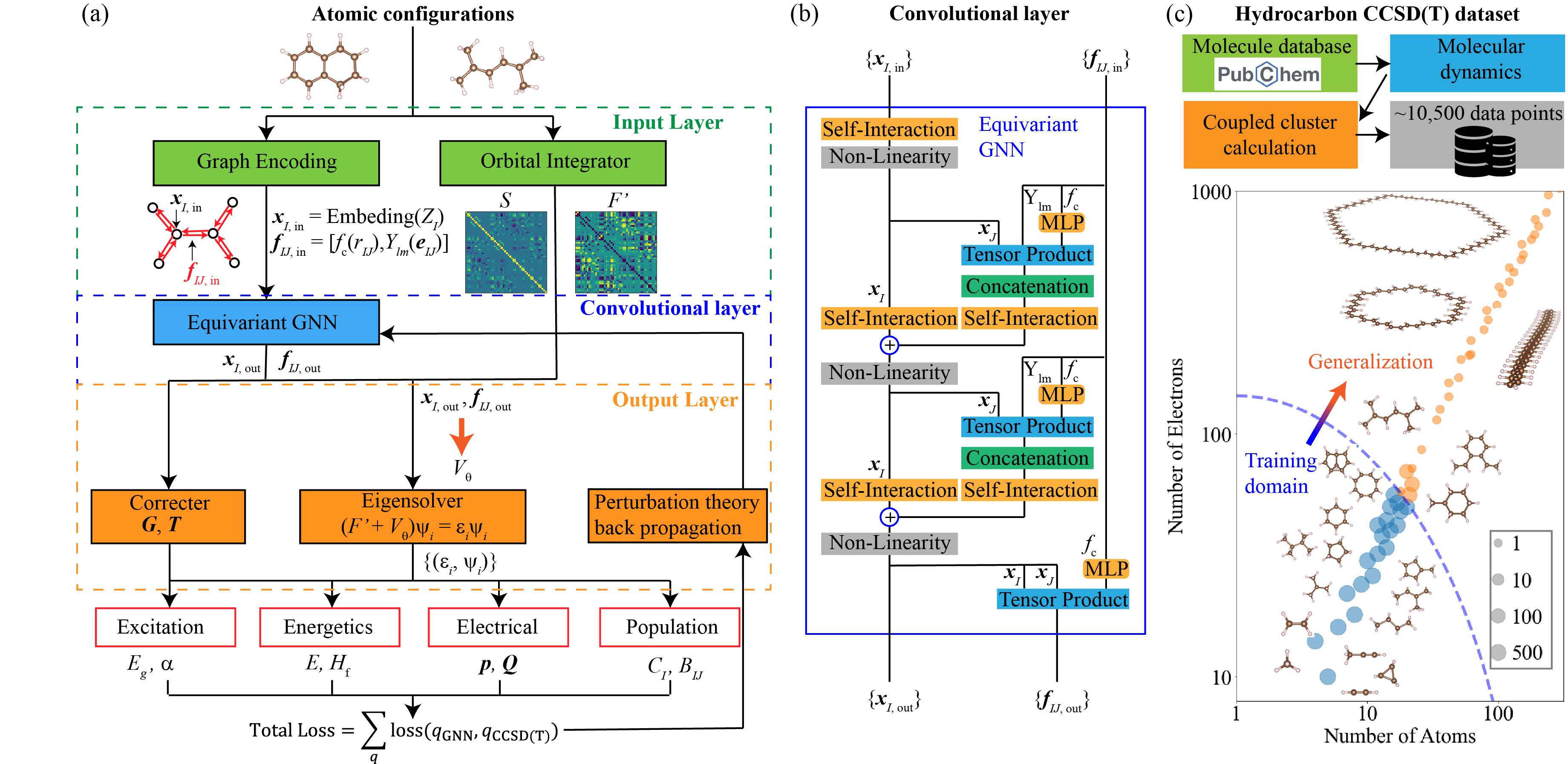}
\caption{Schematic of the EGNN electronic structure workflow. (a) Computation graph of the EGNN method that calculate multiple quantum chemical properties from atomic configurations inputs. The computational graph consists of input layer (green blocks), convolutional layer (blue block), and output layer (orange blocks). (b) Model architecture of the EGNN that consists of two layers of graph convolution and output both node feature $\mathbf x_{I, \rm out}$ and edge feature $\mathbf f_{IJ, \rm out}$. (c) Training and testing dataset generation. About 10,500 atomic configurations of 85 different hydrocarbon molecules are sampled from molecular dynamics trajectories. Data points are plot in the map of number of electrons and atoms, and the dot size reflects the number of training data with the same chemical formula.}
\label{fig:illustration}
\end{figure*}

In comparison, the correlated wavefunction method CCSD(T) is considered the gold-standard in quantum chemistry~\cite{bartlett2007coupled}. It provides high accuracy predictions on various molecular properties. Unfortunately, the computational cost of CCSD(T) calculations has a rather unfavorable scaling relationship with system size. Hence, it can only handle small molecules with up to hundreds of electrons. This urges the combination of CCSD(T) and ML methods, which can potentially have both high accuracy and low computational cost. However, above-mentioned ML methods that directly fit the single-configurational representation of the DFT calculations cannot be applied  with the CCSD(T) training data. This is because CCSD(T) does not provide either a Kohn-Sham (KS) Hamiltonian or single-body molecular orbitals due to the many-body quantum entanglement nature of its representation. 

In this work, we develop a unified multi-task ML method for molecular electronic structures. Instead of focusing solely on energy, our method also provides accurate predictions for various electronic properties; compared with ML models trained on DFT datasets, our method learns from CCSD(T)-accuracy training data. The method incorporates the equivariant graph neural network (EGNN)~\cite{TakamotoIL22,TakamotoOLL23,batzner20223,geiger2022e3nn}, where vectors and tensors are involved in the message-passing step. Using hydrocarbon organic molecules as a testbed, our method predicts molecular energy within chemical accuracy as compared with both CCSD(T) calculation and experiments, and predicts various properties, including electric dipole and quadrupole moments, atomic charge, bond order, energy gap, and electric polarizability with better accuracy than B3LYP, one of the most widely used hybrid DFT functional~\cite{tirado2008performance}. Our trained model shows robust generalization capability from small molecules in the training dataset (molecular weight $< 100$) to larger molecules and even semiconducting polymers (molecular weight up to several thousands). Systematically predicting multiple electronic properties using a single model with local DFT computational speed, the method provides a high-performance tool for computational chemistry and a promising framework for ML electronic structure calculations.

\section{Results}

\subsection{Theory and Model Architecture}

In this section, we briefly describe the theoretical background and model architecture of our EGNN method. 
Basically, we use a NN to simulate the non-local exchange-correlation interactions of a many-body system. Then, a physics-informed approach is used to predict multiple properties from the output of a single NN.

\noindent \textbf{Computational Workflow.} Given an input atomic configuration, our methods output an effective single-body Hamiltonian matrix that predicts quantum chemical properties, as shown in Fig.~\ref{fig:illustration}a. The workflow consists of the input layer, the convolutional layer, and the output layer.

The input layer takes atomic configurations as input, including the information of atomic numbers $(Z_1, Z_2, \cdots ,Z_n)$ and atomic coordinates $(\vec{r}_1, \vec{r}_2,\cdots ,\vec{r}_n)$ of a $n$-atom system. A molecular graph is constructed, where atoms are mapped to graph nodes, while bonds between atoms (neighboring atoms within a cut-off radius $r_{\rm cut} = 2$ \AA ) are mapped to graph edges.  The atomic configuration is then encoded into the node features $\mathbf x_{I, \rm in}$ for atom information and edge features $\mathbf f_{IJ,\rm in}$ for bond information (see Methods A for details). The electron wavefunction is represented using an atomic orbital basis set $\{|\phi_{I,\mu}\rangle\}$~\cite{dunning1989gaussian}, where $I$ is the index of atom and $\mu$ is the index of atomic orbital basis function. Then, a quantum chemistry calculation~\cite{neese2020orca} (the Orbital Integrator block in Fig.~\ref{fig:illustration}a) is used to evaluate single-body effective Hamiltonian $F_{I\mu ,J\nu}$ and the overlap matrix $S_{I\mu ,J\nu}\equiv \langle \phi_{I,\mu}|\phi_{J,\nu}\rangle$ in the non-orthogonal atomic-orbital representation (where $I\mu$ is the row index and $J\nu$ is the colunm index). The $F_{I\mu ,J\nu}$ matrix is obtained by a fast-to-evaluate single-configurational method such as local DFT~\cite{perdew1986density}. 
The Lowdin-symmetrized KS Hamiltonian~\cite{lowdin1950non} is then obtained as
\begin{equation}
    {\bf F}'\equiv {\bf S}^{-1/2}{\bf F}{\bf S}^{-1/2}+\frac{E_{\rm MB}}{n_e}{\bf I},
    \label{eq:F}
\end{equation}
where the last term is an identity shift to account for the many-body energy term (see Methods A for details).  Note that  $\bf{F}'$ is a local DFT-level effective Hamiltonian, meaning that it is easy and fast to compute, but its accuracy is relatively low. We will use $\bf{F}'$ as the starting point of our ML model, and the total effective Hamiltonian of the system ${\bf H}^{\rm eff} = {\bf F}' + {\bf V}^\theta$ is obtained by adding the ML correction term ${\bf V}^\theta$. As  ${\bf F}'$ is obtained from a local DFT calculation, it only contains local exchange-correlation contribution, and the correction term ${\bf V}^\theta$ would account for the non-local exchange-correlation effects. Generally, the non-local exchange-correlation effects can be incorporated in CCSD(T) calculations. However, as mentioned before, the computational costs of CCSD(T) methods are formidably high for large system. The essence of our ML method is to obtain the non-local exchange-correlation effects from a NN, whose computational cost scales only \emph{linearly} with system size.

To obtain the ML correction term, we build a neural network model to predict  ${\bf V}^\theta$. The EGNN framework is employed for the convolutional layer because of its outstanding performance in predicting molecular properties~\cite{batzner20223}. The convolutional layer transforms the input node features $\mathbf x_{I,\rm in}$ and edge features $\mathbf f_{IJ,\rm in}$ to the output node features $\mathbf x_{I,\rm out}$ and edge features $\mathbf f_{IJ,\rm out}$. The EGNN includes a series of linear transformation (Self-Interaction block), activation function (Non-Linearity block), and graph convolution (Tensor Product and Concatenation blocks) layers, as shown in Fig.~\ref{fig:illustration}b. Details on the numerical form and dimension of each block are elaborated in Methods B. The output $\mathbf f_{IJ,\rm out}$ and $\mathbf x_{I,\rm out}$ encode equivariant features of atoms and bonds as well as their atomic environment.

Then, we construct an equivariant ML correction Hamiltonian $\bf{V}^\theta$ from the output features $\mathbf x_{I,\rm out}$ and $\mathbf f_{IJ,\rm out}$ as follow:  
\begin{equation}
    V_{I\mu ,J\nu}^{\theta} = 
    \begin{cases}
        \left[V_{\rm node}(\mathbf x_{I,\rm out})\right]_{\mu ,\nu} & \text{if $I =J$} \\
        \frac{1}{2}\left[V_{\rm edge}(\mathbf f_{IJ,\rm out})\right]_{\mu ,\nu} + \frac{1}{2}\left[V_{\rm edge}(\mathbf f_{JI,\rm out})\right]_{\nu ,\mu} & \text{if $I \neq J$}
    \end{cases}
    \label{eq:Heffective}
\end{equation}
where $V_{\rm node}(\mathbf x_{I,\rm out})$ is a $N_I\times N_I$ symmetric matrix rearranged from node features  $\mathbf x_{I,\rm out}$, while $V_{\rm edge}(\mathbf f_{IJ,\rm out})$ is a $N_I\times N_J$ matrix obtained from edge features $\mathbf f_{JI,\rm out}$.  Here $N_I, N_J$ are the numbers of basis functions of the atom $I,J$. Note that the output matrices $\bf{V}^\theta$ are Hermitian and equivariant under rotation according to the transformation rule of the basis set $\{|\phi_{I,\mu}\rangle\}$ (see Methods B for details). 

An effective electronic structure of the molecule is obtained by solving the eigenvalue equations of the total Hamiltonian $H_{I\mu ,J\nu}^{\rm eff}$:
\begin{equation}
    \sum_{J,\nu} H_{I\mu ,J\nu}^{\rm eff} c_{J,\nu}^i = \epsilon_i c_{I,\mu}^i,
    \label{eq:schrodinger}
\end{equation}
where $\epsilon_i$ gives the $i$-th energy levels, and $c^i_{I,\mu}$ gives the corresponding molecular orbitals through basis expansion $|\psi_i\rangle = \sum_{I,\mu} \tilde{c}^i_{I,\mu}|\phi_{I,\mu}\rangle$ with $\tilde{c}^i = S^{-1/2}c^i$. 

\noindent \textbf{Multiple Learning Tasks.} Our scheme aims to predict multiple observable molecular properties (more than just energy). In order to achieve reduced computational costs, we do not include information about the entire electronic Hilbert space as learning targets. The energy levels and molecular orbitals are used to evaluate a series of ground state properties $O_g$ according to the rules of quantum mechanics:
\begin{equation}
    O_g^{\rm EGNN} = f_{O_g} (\{\epsilon_i\}, \{\mathbf c^i\}), \quad O_g = E, \vec{p}, \mathbf Q, C_I, B_{IJ},
    \label{eq:ground}
\end{equation}
where properties $O_g$ goes through the ground state energy ($E$), electric dipole ($\vec{p}$) and quadrupole ($\mathbf{Q}$) moments, Mulliken atomic charge~\cite{mulliken1955electronic} of each atom $C_I$, and Mayer bond order~\cite{mayer2007bond} of each pair of atoms $B_{IJ}$.  We also evaluate the energy gap (first excitation energy, $E_{\rm g}$) and static electric polarizability $\alpha$:
\begin{equation}
\begin{aligned}
    E_g^{\rm EGNN} & = f_{E_g} (\{\epsilon_i\}, \{\mathbf c^i\}, \mathbf G), \\
    \alpha^{\rm EGNN} & = f_{\alpha} (\{\epsilon_i\}, \{\mathbf c^i\}, \mathbf T).
    \end{aligned}
    \label{eq:excited}
\end{equation}
In principle, the ground state electronic structure does not contain the information of the energy gap and electric polarizability. Therefore, we use the EGNN-output correction terms $\mathbf{G}$ (energy gap correction) and $\mathbf{T}$ (dielectric screening matrix) to account for the excited states information and the linear response information, respectively. The function forms of $f_{O_g}$, $f_{E_g}$, and $f_\alpha$ are elaborated in Methods C. Note that these properties are all derived from the underlying electronic structure, so they are internally related. Therefore, multi-task learning methods can utilize these relations to enhance the model's generalization capability.

\begin{figure*}[t]
\centering
\includegraphics[width=\linewidth]{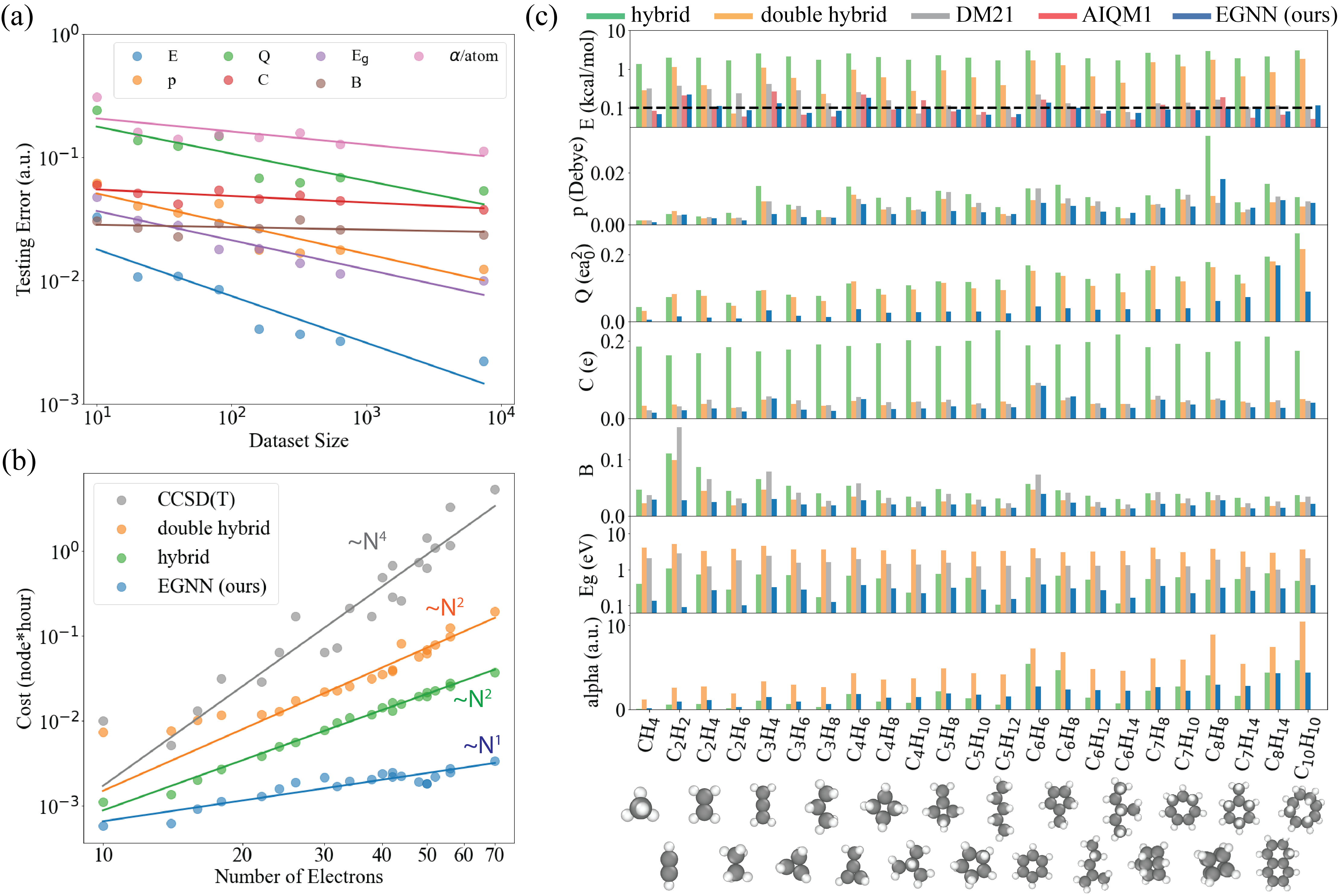}
\caption{Benchmark of the model performance on the testing dataset. (a) Testing root-mean-square errors (RMSE) of different quantities as a function of training dataset size. (b) Computational costs of different methods plot against number of electrons. The computational costs is measured as the calculation time (node hour) on a single Intel Xeon Platinum 8260 CPU node with 48 cores on the MIT SuperCloud~\cite{reuther2018interactive} with sufficient memory for all calculations. The scaling deviates from the theoretical asymptotic scaling, e.g., $N^7$ for CCSD(T),  because the parallelization efficiency is higher for larger molecules. In principle, the $N^7$ scaling for CCSD(T) would appear in the large $N$ limit. (c) Prediction RMSE of the energy ($E$ per atom, reference to separate atoms), electric dipole moment ($\vec{p}$), electric quadrupole moment ($\bf{Q}$), Mulliken atomic charge ($C$), Mayer bond order ($B$), energy gap (1$^{\rm st}$ excitation energy, $E_{\rm g}$), and static electric polarizability ($\alpha$, a.u. means atomic unit). Our EGNN method is compared with the B3LYP hybrid functional, DSD-PBEP86 double hybrid functional~\cite{kozuch2013spin}, DM21 ML functional~\cite{kirkpatrick2021pushing}, and AIQM1 ML potential~\cite{zheng2021artificial, dral2024mlatom}. A representative atomic configuration of each chemical formula is plotted for illustration.}
\label{fig:benchmark}
\end{figure*}

\begin{table*}
\renewcommand\arraystretch{1.4}
\caption{Benchmark of EGNN model's RMSE in predicting different quantum chemical properties on the in-domain (ID) testing dataset and out-of-domain (OOD) validation dataset (written as ID RMSE/OOD RMSE in the table).}
\begin{tabular}{ p{3.6cm}p{1.8cm}p{2.3cm}p{2.3cm}p{2.3cm}p{2.3cm}p{2.3cm}  }
 \hline
 \hline
\bf{RMSE (ID/OOD)} & Unit & Hybrid & Double hybrid & DM21 & AIQM1 & EGNN (ours) \\ 
 \hline 
Energy (per atom) & kcal/mol & 2.20/2.41&0.94/1.20&0.22/0.11&0.13/0.06&\bf 0.11/0.10\\ 
 \hline 
Electric Dipole & 10$^{-2}$ Debye & 1.27/1.20&0.71/0.70&0.78/0.88& -- &\bf0.63/0.81\\ 
 \hline 
Electric Quadrupole & ea$_0^2$ & 0.12/0.21&0.11/0.18& -- & -- &\bf0.03/0.12\\ 
 \hline 
Mulliken Atomic Charge & e & 0.19/0.20&0.04/0.05&0.05/0.04& -- &\bf0.04/0.03\\ 
 \hline 
Mayer Bond Order & -- & 0.05/0.03&0.04/0.02&0.06/0.03& -- &\bf0.02/0.02\\ 
 \hline 
1$^{st}$-Excitation Energy & eV & 0.59/0.63&3.71/3.26&1.71/1.47& -- &\bf0.26/0.31\\ 
 \hline 
Polarizability & a.u. & 2.22/4.32&4.74/8.05& -- & -- &\bf1.85/3.91\\ 
 \hline 
 \hline
\end{tabular}
\label{table:complexity}
\end{table*}

The multi-task learning is implemented by minimizing the total loss function $L_{\rm Total}$ constructed as follows:
\begin{equation}
\begin{aligned}
    L_{\rm Total} &=  \sum_{O\in \{O_g, E_g, \alpha\}} l_{O} + l_V, \\ 
    l_{O} &= w_{O}\times {\rm MSEloss}(O^{\rm EGNN}, O^{\rm label}), \\
    l_V &=\frac{w_V}{N_{\rm basis}^2}\sum_{I\mu ,J\nu}\vert V^\theta_{I\mu ,J\nu}\vert^2.
\end{aligned}
\end{equation}
Here for each property $O$, $l_O$ is the the mean-square error (MSE) loss between $O^{\rm EGNN}$ and $O^{\rm label}$, the EGNN predictions (Eq.~\eqref{eq:ground}\eqref{eq:excited}) and coupled-cluster labels in the training dataset, respectively. Meanwhile, $l_V$ is a regularization that penalizes large correction matrix $\bf{V}_\theta$, and $N_{\rm basis}$ is the total number of basis functions in the molecule. The weights $w_V$ and $w_O$ are hyperparameters whose values are listed in supplementary information (SI) section I. The weights are chosen to balance the training tasks so that the training errors of all tasks decrease to satisfactory levels. Minimizing $L_{\rm Total}$ requires the back-propagation through Eq.~\eqref{eq:schrodinger} (i.e., calculating $\partial \epsilon_i/\partial \mathbf H^{\rm eff}$ and $\partial \mathbf c^i/\partial \mathbf H^{\rm eff}$), which is numerically unstable~\cite{eigh}. To overcome this issue, we derive customized back-propagation schemes for each property using perturbation theory in quantum mechanics (see SI section S1 for details), giving
\begin{equation}
\begin{aligned}
    \nabla_\theta \epsilon_i  &= (\mathbf c^i)^\dagger  (\nabla_\theta \bf{V}^\theta ) \mathbf c^i\\
    \nabla_\theta \mathbf{c}^i &= \sum_{p\neq i} \frac{(\mathbf{c}^p)^\dagger (\nabla_\theta \mathbf{V}^\theta ) \mathbf{c}^i}{\epsilon_i - \epsilon_p}\mathbf c^p.
    \label{eq:grad}
\end{aligned}
\end{equation}
When evaluating the gradients of properties in Eqs.~\eqref{eq:ground} and \eqref{eq:excited} using the chain rule, terms that analytically cancel each other are removed in numerical evaluation, making the scheme numerically stable.

\noindent \textbf{Software Realization.} Atomic configurations of molecules in our training dataset are generated by the workflow shown in Fig.~\ref{fig:illustration}c. First, 85 hydrocarbon molecule structures are collected from the PubChem database~\cite{pubchem}. Molecular dynamics (MD) simulation with TeaNet interatomic potential~\cite{TakamotoIL22,TakamotoOLL23} is then performed for each molecule structure to sample an ensemble of atomic configurations. Then, the CCSD(T) calculations are used to calculate the ground-state properties of sampled configurations, and the EOM-CCSD~\cite{krylov2008equation} calculations are used to calculate the excited-state properties. The $\bf{S}$ and $\bf{F}$ matrices are evaluated by the ORCA quantum chemistry program package~\cite{neese2020orca} with the fast-to-evaluate BP86 local density functional~\cite{perdew1986density}  and the medium-sized cc-pVDZ basis set~\cite{dunning1989gaussian}. The EGNN model based on the e3nn package~\cite{geiger2022e3nn}  is trained on small-molecules training dataset (training domain, Fig.~\ref{fig:illustration}c) and tested on both small molecules in the training dataset (in-domain validation) and larger molecules outside the training dataset (out-of-domain validation). Details about training and testing dataset and training hyperparameters are elaborated in SI section S2. 

\subsection{Model Performance and Applications}

In this section, we benchmark the performance of our EGNN model and display potential applications of the model in systems of practical importance. 
We will make direct comparisons with experimental results wherever possible, and the results suggest the outstanding generalization capability and predicting power of our approach.

\noindent \textbf{Benchmark of Model Performance.} The model's generalization capability from small molecules to large molecules is essential for its usefulness on complex systems where coupled cluster calculations cannot be implemented on current computational platforms, due to their formidable computational costs. To test the generalization capability and data efficiency of our model, we train the model with varied training dataset size $N_{\rm train}$, ranging from 10 to 7440 atomic configurations. The testing root-mean-square error (RMSE, absolute error in atomic units) of different trained properties exhibit a decreasing trend when the training dataset size increases, as shown in Fig.~\ref{fig:benchmark}a, indicating effective model generalization. Notably, the energy error has the fastest drops with a slope of $-0.38$ (that means the testing error $\propto N_{\rm train}^{\rm -0.38}$). In comparison, some of the recently developed advanced ML potentials (that directly learn energies and their derivatives) exhibits lower slopes of about -0.25~\cite{batzner20223, merchant2023scaling}. This implies potential advantage of our multi-task method: as our multi-task method learns different molecular properties through a shared representation (the electronic structure), the domain information learned from one property can help the model's generalization on predicting other properties~\cite{caruana1997multitask}, providing outstanding data efficiency. 

\begin{figure*}[t]
\centering
\includegraphics[width=\linewidth]{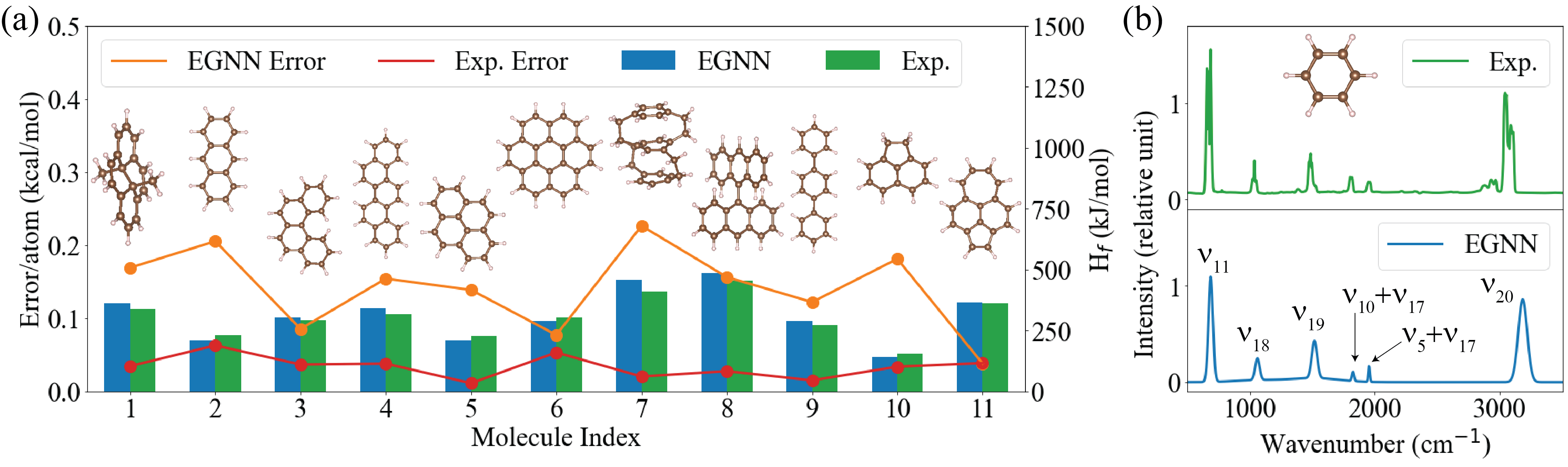}
\caption{Validation of the EGNN's predictions on gas phase aromatic hydrocarbon molecules, as compared with experimental results. (a) Standard enthalpy of formation. The EGNN predictions and experimental values from Ref.~\cite{slayden2001energetics} (right axis) are compared for 11 molecules (see SI Table II for details). The difference between the EGNN method and experimental values are shown by the orange line, and the experimental uncertainty is shown by the red line (left axis). (b) Infrared spectrum of benzene. The experimental data is from the NIST Chemistry WebBook~\cite{NISTwebbook}. Vibration modes corresponding to the peaks are labeled following the convention in Ref.~\cite{wilson1934normal}.}
\label{fig:exp}
\end{figure*}

Then, we benchmark the computational costs and prediction accuracy of our model trained on 7440 atomic configurations with 70 different molecules, which will be used in the rest of this paper. Our EGNN method exhibits significantly smaller computational cost and slower scaling with system size, as compared with the hybrid functional, double hybrid functional~\cite{kozuch2013spin}, and CCSD(T) method (Fig.~\ref{fig:benchmark}b). Compared to the hybrid functional, our method avoids the expensive calculation of the exact exchange thus requiring much smaller computational costs~\cite{tirado2008performance}. Using the gold-standard CCSD(T) calculation as a reference, the prediction accuracy of our EGNN method on various molecular properties is compared with that of the several theory functionals and existing ML methods, as shown in Fig.~\ref{fig:benchmark}c and Table~\ref{table:complexity}. The comparison is implemented on both an in-domain (ID) and an out-of-domain (OOD) testing dataset of hydrocarbon molecules.
The EGNN predictions exhibit smaller RMSE than the hybrid, double hybrid, and DM21~\cite{kirkpatrick2021pushing} functional on most compared molecular properties (except electric dipole moment on the OOD dataset, where double hybrid gives the smallest RMSE). Remarkably, the RMSE of the combination energy predicted by the EGNN is about 0.1 kcal/mol (about 4 meV) per atom in both the ID and OOD dataset. Our method exhibits similar RMSE of combination energy compared to the AIQM1 ML potential that features its chemical accuracy energy predictions. These results confirm that our EGNN's predictions on reaction energies can approach the quantum chemical accuracy (assuming that on average 1 mol molecules in reactants contain $\sim$10 mol atoms). Note that besides the ground-state properties, our EGNN also provides excited-state property $E_{\rm g}$ and linear response property $\alpha$ with better overall accuracy than other methods in comparison (Table \ref{table:complexity}), though there are certain molecules where the hybrid functional performs better. The statistical distribution of the prediction errors of the EGNN and B3LYP hybrid functional are shown in Fig.~S1 in SI. Although the hybrid functional gives smaller median error for $\alpha$, the EGNN is less likely to make large errors (more robust), leading to  better root-mean-square error. 

\noindent \textbf{Aromatic Molecules.} Hydrocarbon molecules have a vast structural space, including various types of local atomic environments (see a few examples below Fig.~\ref{fig:benchmark}c). Our EGNN method provides a single model that exhibits generalization capability among the vast hydrocarbon structural space (cf. Fig.~\ref{fig:benchmark}). To further examine the model's generalization capability in more complex structures, we apply the EGNN model to a series of aromatic hydrocarbon molecules synthesized in experiments~\cite{slayden2001energetics}. The gas phase standard enthalpy of formation $H_{\rm f}$ is an essential thermochemical property of molecules that can be accurately measured in experiments. In this regard, we use the EGNN model to predict $H_{\rm f}$ of various aromatic molecules in a comprehensive experimental review paper Ref.~\cite{slayden2001energetics}.  Among the 71 molecules lists in Ref.~\cite{slayden2001energetics},  we calculate $H_{\rm f}$ of those molecules with serial numbers (defined in Ref.~\cite{slayden2001energetics}) dividable by 4 and compare them with experiments, as shown in Fig.~\ref{fig:exp}a. The selected molecules cover various classes of aromatic molecules, including polycyclic aromatic hydrocarbons, cyclophanes, polyphenyl, and nonalternant hydrocarbons. The EGNN predictions on $H_{\rm f}$ are well consistent with experiments on all molecules, and their difference is only around $0.1\sim 0.2$ kcal/mol per atom. Note that the EGNN prediction error is on the same order of magnitude as the experimental error bar (though numerically larger), indicating high prediction accuracy.

Besides thermochemical properties, our EGNN model can also predict spectral properties, as shown in Fig.~\ref{fig:exp}b. Especially, infrared (IR) spectrum reflects essential information of molecular vibrational modes and their interaction with light. In previous work of ML electronic structure~\cite{shao2023machine}, although the predicted peak positions of the IR spectrum are well consistent with the experiment, the predicted peak intensity are inconsistent with the experiment. In comparison, our EGNN model predicts both the peak positions and intensity well consistent with the experiment, providing both the fundamental bands and combination bands known as ``benzene fingers" in the IR spectrum. The good consistency of peak intensity is attributed to accurate predictions on the transition dipole moments that determine the intensity of light-matter interaction. Details on calculating the IR spectrum is elaborated in SI section S3.

\begin{figure}[t]
\centering
\includegraphics[width=\linewidth]{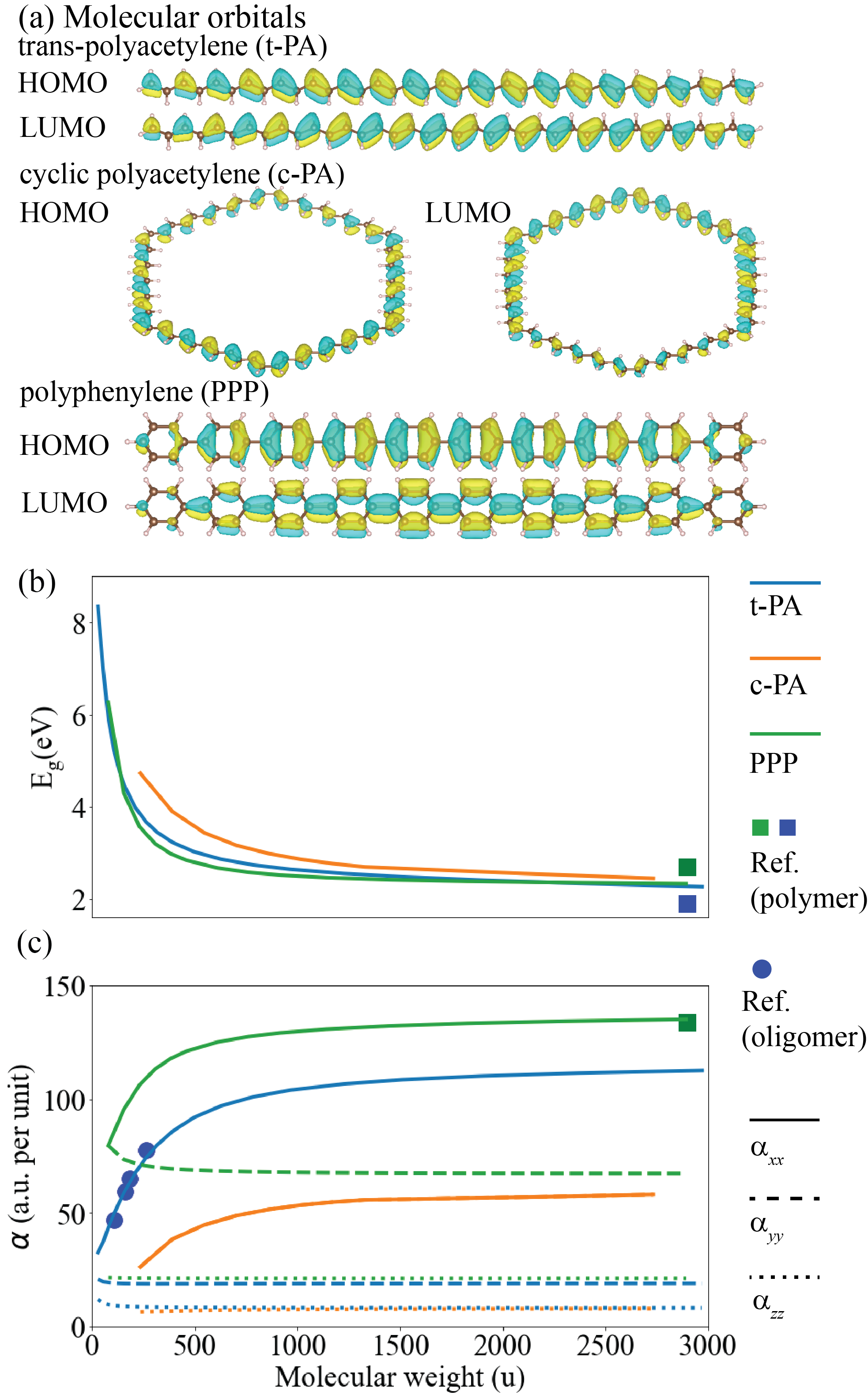}
\caption{EGNN predictions for the electronic proerpties of semiconducting polymers. (a) Atomic structure and HOMO wavefunctions of t-PA, polyphenylene PPP, and c-PA. The HOMO wavefunctions are visualized by isosurfaces at the level of $\pm 0.01$ \AA$^{-2/3}$ (positive isosurface colored blue and negative isosurface colored yellow). (b) Energy gap and (c) static electric polarizability of t-PA (blue lines), PPP (green lines), and c-PA (orange lines) with different polymer chain length. Longitudinal polarizability $\alpha_{xx}$, horizontal polarizability $\alpha_{yy}$, and vertical polarizability $\alpha_{zz}$ are shown as solid, dashed, and dotted lines, respectively. Squares (blue for t-PA and green for PPP) represent literature values for polymers in experiments~\cite{grem1992realization, heeger2001nobel} and correlated calculations~\cite{otto2004dynamic}, and blue dots represent literature values for t-PA oligomers from the MP2 correlated calculations~\cite{champagne1998assessment}.}
\label{fig:polymer}
\end{figure}

\noindent \textbf{Large-scale Semiconducting Polymers.} Besides small molecules, we also apply the EGNN model to semiconducting polymers consisting of hundreds of atoms, which are difficult to calculate by rigorous correlated methods such as CCSD(T). Semiconducting polymers are organic macromolecules with small energy gap and high electrical conductivity compared to insulating polymers. Because of these electronic features, semiconducting polymers attract broad research interests both for the fundamental understanding of quantum chemistry~\cite{heeger2001nobel} and for applications in semiconductor industry~\cite{geffroy2006organic}. The essential electronic properties of semiconducting polymers originates from the $\pi$-bonds with delocalized molecular orbitals. As the delocalized molecular orbitals extend through the whole macroscopic molecule (Fig.~\ref{fig:polymer}a), the polymers' electronic properties also involve long-range correlation, making it challenging for ML methods. Therefore, it is important to examine whether the EGNN model can capture semiconducting polymers' electronic properties involving delocalized molecular orbitals.

Three kinds of semiconducting polymers, trans-polyacetylene (t-PA), cyclic polyacetylene (c-PA), and polyphenylene (PPP), are studied using our EGNN model. The model correctly captures the delocalized $\pi$-bond feature of frontier orbitals (highest occupied molecular orbital (HOMO) and lowest unoccupied molecular orbital (LUMO)), as shown in Fig.~\ref{fig:polymer}a. In the t-PA, the HOMO and LUMO are on the carbon-carbon double and single bonds, respectively, consistent with the predictions from the renowned  Su–Schrieffer–Heeger (SSH) model~\cite{heeger2001nobel}. In the c-PA, the left/right mirror symmetry of the molecule is correctly reflected in the predicted frontier orbitals, which exhibits odd parity upon mirror reflection. Both the HOMO and LUMO have two anti-phase boundaries where the wavefunctions shift from double bonds to single bonds. The same symmetry is also reflected in the PPP with odd-parity HOMO and even-parity LUMO. It should be emphasized that the molecular orbitals are not included as fitting targets in our model training. However, as the model learned relevant properties such as electric moments, atomic charge, and bond order, it gives at least qualitatively correct predictions for frontier orbitals in these semiconducting polymers.

Various important electronic properties of semiconducting polymers depend on the chain length, including the energy gap $E_{\rm g}$ and polarizability $\alpha$. We calculate such chain-length dependence (up to more than 400 atoms) using our EGNN method, as shown in Fig.~\ref{fig:polymer}b,c. One can see that $E_{\rm g}$ is larger for oligomers with shorter chain length due to the quantum confinement effect and converges to a smaller value for long polymer. The converged energy gap for long t-PA and PPP polymers calculated by our EGNN are in reasonable agreement with the experimental values~\cite{heeger2001nobel, grem1992realization} (relative errors within 10\%), which are shown as squares in Fig.~\ref{fig:polymer}b. The longitudinal static electric polarizability $\alpha_{xx}$ (per monomer) is positively related to the polymer chain length. This is because in longer chains, more delocalized electron distributions can have larger displacements under external electric field. The predicted $\alpha_{xx}$ for t-PA oligomers and PPP polymers are in perfect agreement with previous correlated calculations using the high-accuracy MP2 method~\cite{otto2004dynamic,champagne1998assessment}, as shown in Fig.~\ref{fig:polymer}c. The chain length-dependent $E_{\rm g}$ and $\alpha$ of cyclic PA, to the best of our knowledge, have not been reported. We provide their values as a prediction to be examined by future work.

\section{Conclusion and Outlook}

In this work, we developed a multi-task learning method for predicting molecular electronic structure with coupled-cluster-level accuracy. In our method, an EGNN is trained on a series of molecular properties in order to capture their shared underlying representation, the effective electronic Hamiltonian. Our EGNN model shows significantly lower computational costs and higher prediction accuracy on various molecular properties than the widely-used B3LYP functional. 

Compared to other works that take the electronic structure as a direct fitting target~\cite{schutt2019unifying,unke2021se,li2022deep, gong2023general}, our method takes the electronic structure as a shared representation to predict various molecular properties. This approach, on the one hand, enables training on data beyond the DFT accuracy; on the other hand, relieve the burden of fitting all matrix elements of the electronic Hamiltonian, which are not direct quantum observable. The later enables us to go beyond the minimal basis used in previous works by a relatively small NN with only 511,589 parameters. With the electronic structure as a physics-informed representation, our method exhibits generalization capability for challenging systems with delocalized molecular orbitals. In such systems, atomic structure at one position has long-range influence to electronic features (such as electron density) at another distant position. Although the direct output of our EGNN, the matrix elements of $\bf{V}_\theta$, only depends on local atomic environments, the long-range influence is captured through the output-layer calculations based on the rules of quantum mechanics. In comparison, such long-range influence is unlikely to be captured by directly using graph neural network or local atomic descriptors to predict molecular properties without a physics-informed representation.

Although the current EGNN model is limited to hydrocarbon molecules, our method can be readily applied to systems with different elements. Applying the method to datasets with more elements can produce a general-purpose electronic structure predictor, which is left to future work.

\section{Methods}
\subsection{Graph encoding of atomic configuration}
The atomic numbers $Z_I$ of input elements are encoded as node features $\mathbf{x}_{I,\rm in}$ by one-hot embedding. In our case, hydrogen and carbon atoms are encoded as scalar arrays $[0,1]$ and $[1,0]$, respectively. The atomic coordinates are encoded as edge features $\mathbf{f}_{IJ,\rm in}\equiv [f_c(r_{IJ}), Y_{lm}(\vec{e}_{IJ})]$, where $f_c(r) \equiv \frac{1}{2}\left[\cos{(\pi\frac{r}{r_{\rm cut}})}+1\right]$ is a smooth cut-off function reflecting the bond length $r_{IJ}\equiv |\vec{r}_I-\vec{r}_J|$~\cite{zhang2018end}, and $Y_{lm}(\vec{e}_{IJ})$ is the spherical harmonic functions acting on the unit vector $\vec{e}_{IJ}\equiv \frac{\vec{r}_I-\vec{r}_J}{|\vec{r}_I-\vec{r}_J|}$ representing the bond orientation~\cite{geiger2022e3nn}. We include $Y_{lm}$ tensors up to $l=2$. 

As the hydrocarbon molecules we study are all close-shell molecules, we use spin-restricted DFT calculations to obtain ${\bf F}$ and we assume ${\bf V}_\theta$ is spin-independent throughout this paper. Namely, the spin-up and spin-down molecular orbitals and energy levels are the same, and all molecular orbitals are either doubly occupied or vacant.
The electronic energy of the BP86 DFT calculation $E_{\rm BP86}$ equals the band structure energy $2\sum_{i=1}^{n_e/2} \epsilon_i$ (where $\epsilon_i$ is the $i$-th molecular orbital energy level and $n_e$ is the number of electrons) plus a many-body energy $E_{\rm MB}$:
\begin{equation}
    E_{\rm BP86} = 2\sum_{i=1}^{n_e/2} \epsilon_i + E_{\rm MB}
\end{equation}
$E_{\rm MB}$ originates from the double-counting of electron-electron interaction in the band structure energy and is obtained from the output of the ORCA BP86 DFT calculation. In order to incorporate the many-body energy into the KS effective Hamiltonian ${\bf F}$, we construct ${\bf F}'$ as in Eq.~\eqref{eq:F}, so the direct summation of band structure energies given by $F$ equals:
\begin{equation}
\begin{aligned}
    2\sum_{i=1}^{n_e/2} {\rm eig}_i({\bf F}') &= 2\sum_{i=1}^{n_e/2} {\rm eig}_i({\bf S}^{-1/2}{\bf F}{\bf S}^{-1/2}+\frac{E_{\rm MB}}{n_e}{\bf I})\\
    &=2\sum_{i=1}^{n_e/2} \left[{\rm eig}_i({\bf S}^{-1/2}{\bf FS}^{-1/2})+\frac{E_{\rm MB}}{n_e}\right] \\
    &= 2\sum_{i=1}^{n_e/2} \epsilon_i + E_{\rm MB},
\end{aligned}
\end{equation}
where ${\rm eig}_i$ is a function that returns the $i$-th lowest eigenvalue of a matrix, and we use the fact that the energy level $\epsilon_i$ is the eigenvalue of the Lowdin-symmetrized Hamiltonian ${\rm eig}_i(S^{-1/2}FS^{-1/2})$~\cite{lowdin1950non}. After this transformation, the KS effective Hamiltonian (both before and after the $\bf{V}_\theta$ correction) already includes the many-body energy, and the total electronic energy is just the summation of band structure energies. Adding the $E_{\rm MB}$ term does not change the eigenfunction and relative energy levels, so that all other properties are not changed by adding the $E_{\rm MB}$ term.

\subsection{Architecture of the convolutional layer}
In the following technical description, we use terminologies defined in the e3nn documentation. One can refer to Ref.~\cite{geiger2022e3nn} for further information. In the convolutional layer, the input feature first goes through a $N_{\rm species}\times N_{\rm species}$ linear transformation (the first Self-Interaction block, $N_{\rm species}$ is the number of different elements in the system) and an activation layer (the first non-Linearity block). All activation layers in our EGNN are realized by tanh function acting on scalar features. 

Then, the input features go through the first-step convolution: a fully connected tensor product (the first Tensor Product block) of node feature $\mathbf x_J$ and the spherical Harmonic components of all connected edge feature $\mathbf f_{IJ}$ mapping to an irreducible representation "8x0e + 8x1o + 8x2e" (denoted as Irreps1), meaning 8 even scalar, 8 odd vector, and 8 even rank-2 tensor. Weights in the fully connected tensor product are from a multilayer perceptron (the first MLP block) taking $f_c(r_{IJ})$ as input. All MLP blocks in Fig.~\ref{fig:illustration}b has a $1\times 16\times 16\times 16\times N_{w}$ structure and tanh activation function, where $N_w$ is the number of weights in the tensor product. Then, in the Concatenation block, tensor products from different edges $\mathbf f_{IJ}$ connected to the node $I$ are summed to a new node feature on $I$. The new node features then go through a linear transformation (Self-Interaction block) that outputs the same data type Irreps1. In all Self-Interaction layers, linear combinations are only applied to features with the same tensor order. The new node features are added to the original node features undergoes a linear transformation, complete the first-step convolution.

The second step convolution has the same architecture. The only difference is that the second Tensor Product block takes input node features of Irreps1 and output features of "8x0e + 8x0o + 8x1e + 8x1o + 8x2e + 8x2o" (denoted as Irreps2), meaning 8 even and  odd scalar, vector, and tensor, respectively. The output of both Self-Interaction blocks are also Irreps2. 

After another activation function, the node features are output as $\mathbf x_{I,\rm out}$. Another tensor product acting on the node features of two endpoints of each edge is applied to get the output bond feature $\mathbf f_{IJ, \rm out}$, also with a dimension of Irreps2 and weight parameters from the MLP taking $f_c(r_{IJ})$ as input.

Finally, the output features are used to construct the correction matrix $\bf{V}_\theta$ through Eq.~\eqref{eq:Heffective}. $V_{\rm node}(\mathbf x_{I,\rm out})$ first apply a linear layer from input dimension of Irreps2 to output dimension $\text{Irreps}(I)^{\otimes 2}$, where:
\begin{equation}
    {\rm Irreps}(I) = \begin{cases} (2\times 0e + 1\times 1o) & \text{if } I \text{ is H}\\
    (3\times 0e + 2\times 1o + 1\times 2e) & \text{if } I \text{ is C}
    \end{cases}
\end{equation}
The output dimension corresponds to the irreducible representation of the block diagonal terms of the Hamiltonian (as cc-pVDZ basis of hydrogen includes two s orbitals (0e) and one group of p orbitals (1o); while that of carbon includes three s orbitals (0e), two groups of p orbitals (1o), and one group of $d$ orbitals (2e)). The output is then arranged into the $N_I\times N_I$ matrix form, $V_{I, \rm out}$, according to the Wigner-Eckart theorem~\cite{gong2023general}, and symmetrized to obtain $V_{\rm node}(\mathbf x_{I,\rm out}) = \frac{\lambda_V}{2}(V_{I,\rm out}+V_{I,\rm out}^T)$. $\lambda$ is a constant hyperparameter set as 0.2 for our model.
Similarly, the off-diagonal term $V_{\rm edge}(\mathbf f_{IJ,\rm out})$ in Eq.~\eqref{eq:Heffective} applies a linear layer from input dimension of Irreps2 to output dimension Irreps($I,J$) that equals:
\begin{equation}
       {\rm Irreps}(I,J) = {\rm Irreps}(I)\otimes {\rm Irreps}(J)
\end{equation}
The output are then arranged into the $N_I\times N_J$ matrix and multiplied by $\lambda_V$, giving $V_{\rm edge}(\mathbf f_{IJ,\rm out})$. 

In addition, the energy gap correction term $\mathbf G$ is obtained from a $8\times 32\times 3$ MLP that takes the even scalars of $\mathbf x_{I,\rm out}$ as input and output a 3-component scalar array, $\mathbf g_{I; 0,1,2}$, with tanh activation. The first component is for attention pooling:
\begin{equation}
    \mathbf G_K = \sum_I \frac{e^{g_{I,0}}}{\sum_J e^{g_{J,0}}} g_{I,K}, \qquad K = 1,2
\end{equation}
Giving the two-component bandgap correction array $\mathbf G$. The polarizability correction term, the screening matrix $\mathbf T$ is obtained from the edge features $\mathbf f_{IJ,\rm out}$ going through a Irreps2 to $32\times 0e + 1\times 2e$ linear layer, an tanh activation layer, and a $32\times 0e + 1\times 2e$ to $1\times 0e + 1\times 2e$ linear layer. The  $1\times 0e + 1\times 2e$ array is then multiplied by a factor $\lambda_T$ (set as 0.01 in our case) arranged into the symmetric matrix $\mathbf T$'s 6 independent components.

\subsection{Evaluating molecular properties}
The ground state properties in Eq.~\eqref{eq:ground} is evaluated by the electronic structure by the following equations~\cite{mulliken1955electronic,mayer2007bond}:
\begin{equation}
\begin{aligned}
    E^{\rm EGNN} &= E_{\rm NN} + 2 \sum_{i=1}^{n_e/2} \epsilon_i \\
    \vec{p}^{EGNN} &= -2e\sum_{i=1}^{n_e/2} \sum_{I\mu ,J\nu} (\tilde{c}_{I,\mu}^i)^* \tilde{c}_{J,\nu}^i \langle\phi_{I,\mu}|\hat{\vec{r}}|\phi_{J,\nu}\rangle\\
    \mathbf Q^{\rm EGNN} &= 2e^2\sum_{i=1}^{n_e/2} \sum_{I\mu ,J\nu} (\tilde{c}_{I,\mu}^i)^* \tilde{c}_{J,\nu}^i \langle\phi_{I,\mu}|\hat{\vec{r}}\hat{\vec{r}}|\phi_{J,\nu}\rangle \\
    C_I^{\rm EGNN} &= e\left[ Z_I - 2\sum_{i=1}^{n_e/2}\sum_{J\mu\nu} (\tilde{c}_{I,\mu}^i)^* \tilde{c}_{J,\nu}^i S_{I\mu,J\nu}\right] \\
    B_{IJ}^{\rm EGNN} &=  4\sum_{i,j=1}^{n_e/2}\sum_{KL\mu\nu\lambda\sigma} (\tilde{c}_{K,\lambda}^i)^* \tilde{c}_{I,\mu}^i S_{K\lambda,J\nu}(\tilde{c}_{L,\sigma}^j)^* \tilde{c}_{J,\nu}^j S_{L\sigma,I\mu}
    \label{eq:groundstate}
\end{aligned}
\end{equation}
where $E_{\rm NN}$ is the Coulomb repulsion energy between nuclei and nuclei,  and $e$ and $\hat{\vec{r}}$ are the electron charge and position operator, respectively.

Besides,  Using the ground state electronic structure obtained from Eq.~\eqref{eq:schrodinger}, $E_{\rm g}$ can be roughly estimated as $\epsilon_{n_e/2+1}-\epsilon_{n_e/2}$, the energy difference between the HOMO and LUMO (abbreviated as the HOMO-LUMO gap). However, in principle, the ground state electronic structure ($\epsilon_n, \mathbf c^n$) does not contain the information of excited states (once a electron is excited, $\epsilon_n$ and $\mathbf c^n$ undergo relaxation and become different). Therefore, we use the EGNN to output two correction terms $G_1$ and $G_2$. $E_{\rm g}$ is then evaluated as a linear transformation of the HOMO-LUMO gap using $G_1$ and $G_2$ as the coefficients:
\begin{equation}
    E_{\rm g}^{\rm EGNN} = (1+G_1)(\epsilon_{n_e/2+1}-\epsilon_{n_e/2})+G_2
    \label{eq:Eg}
\end{equation}
Evaluation of the static electric polarizability is done in two steps. First, we evaluate the single-particle polarizability $\alpha_0$ using perturbation theory:
\begin{equation}
    \alpha_0 = 2e^2\sum_{a=n_e/2+1}^{N_{\rm basis}}\sum_{i = 1}^{n_e/2}\frac{\vec{r}_{ai}\vec{r}_{ia}}{\epsilon_a - \epsilon_i}
    \label{eq:polar}
\end{equation}
where $N_{\rm basis}$ is the number of basis functions of the molecule, and $\vec{r}_{ai}\equiv \sum_{I\mu,J\nu}(\tilde{c}_{I,\mu}^a)^* \tilde{c}_{J,\nu}^i\langle\phi_{I,\mu}|\hat{\vec{r}}|\phi_{J,\nu}\rangle$. However, the single-particle approximation used in Eq.~\eqref{eq:polar} does not consider the electric screening effect from electron-electron interaction. We use the EGNN to output a screening matrix $T$ and evaluate the corrected polarizability $\alpha$ as follow:
\begin{equation}
\alpha^{\rm EGNN} = ({\bf I} + \alpha_0{\bf T})^{-1}\alpha_0.
\label{eq:screened}
\end{equation}

We evaluate the gas phase standard enthalpy of formation of molecules in Fig.~\ref{fig:exp} using atomic configurations built by Avogadro~\cite{hanwell2012avogadro} and relaxed by the BP86 functional with cc-pVDZ basis set. The total energy at the relaxed atomic configuration is then calculated by our EGNN model. The zero-point energy (ZPE) and thermal vibration, rotation, and translation energy at $T=298.15$ K are also calculated by the BP86 functional with cc-pVDZ basis set implemented in ORCA. The ZPE is corrected by the optimal scaling factor of 1.0393 according to Ref.~\cite{kesharwani2015frequency}. Summing all energy terms give the inner energy $U$, and the enthalpy is evaluated as $H \simeq U + Nk_{\rm B}T$ ($N$ is the number of molecules and $k_{\rm B}$ is the Boltzmann constant), where we use the ideal gas law. To obtain the standard enthalpy of formation, we subtract the reference state enthalpy of graphite and hydrogen gas at standard condition. The reference enthalpy for each carbon and hydrogen atom are determined as -38.04639 a.u. and -0.57550 a.u., respectively, using CCSD(T) calculation with cc-pVTZ basis set combined with measured standard enthalpy of formation of atomic carbon, atomic hydrogen, and benzene. Atomic configurations of semiconducting polymers in Fig.~\ref{fig:polymer} are relaxed using the PreFerred Potential v5.0.0~\cite{TakamotoIL22,TakamotoOLL23}.

\section{Data Availability}
Detailed data of our benchmark test results (Fig.~2, Fig.~S1, Table 1) and the calculation results of aromatic molecules (Fig.~3) and semiconducting polymers (Fig.~4) will be made available through figshare. The training and testing dataset is available upon reasonable requests to the corresponding authors.

\section{Code Availability}
The source code to generate the training dataset, train the EGNN model, and apply the trained EGNN model to hydrocarbon molecules has been deposited into a publicly available GitHub repository https://github.com/htang113/Multi-task-electronic. 

\section{Acknowledgements}
This work was supported by Honda Research Institute (HRI-USA).  H.T. acknowledges support from the Mathworks Engineering Fellowship. The calculations in this work were performed in part on the Matlantis high-speed universal atomistic simulator, the Texas Advanced Computing Center (TACC),  the MIT SuperCloud, and the National Energy Research Scientific Computing (NERSC).

\bibliography{bibliography}

\begin{thebibliography}{58}%
\makeatletter
\providecommand \@ifxundefined [1]{%
 \@ifx{#1\undefined}
}%
\providecommand \@ifnum [1]{%
 \ifnum #1\expandafter \@firstoftwo
 \else \expandafter \@secondoftwo
 \fi
}%
\providecommand \@ifx [1]{%
 \ifx #1\expandafter \@firstoftwo
 \else \expandafter \@secondoftwo
 \fi
}%
\providecommand \natexlab [1]{#1}%
\providecommand \enquote  [1]{``#1''}%
\providecommand \bibnamefont  [1]{#1}%
\providecommand \bibfnamefont [1]{#1}%
\providecommand \citenamefont [1]{#1}%
\providecommand \href@noop [0]{\@secondoftwo}%
\providecommand \href [0]{\begingroup \@sanitize@url \@href}%
\providecommand \@href[1]{\@@startlink{#1}\@@href}%
\providecommand \@@href[1]{\endgroup#1\@@endlink}%
\providecommand \@sanitize@url [0]{\catcode `\\12\catcode `\$12\catcode `\&12\catcode `\#12\catcode `\^12\catcode `\_12\catcode `\%12\relax}%
\providecommand \@@startlink[1]{}%
\providecommand \@@endlink[0]{}%
\providecommand \url  [0]{\begingroup\@sanitize@url \@url }%
\providecommand \@url [1]{\endgroup\@href {#1}{\urlprefix }}%
\providecommand \urlprefix  [0]{URL }%
\providecommand \Eprint [0]{\href }%
\providecommand \doibase [0]{https://doi.org/}%
\providecommand \selectlanguage [0]{\@gobble}%
\providecommand \bibinfo  [0]{\@secondoftwo}%
\providecommand \bibfield  [0]{\@secondoftwo}%
\providecommand \translation [1]{[#1]}%
\providecommand \BibitemOpen [0]{}%
\providecommand \bibitemStop [0]{}%
\providecommand \bibitemNoStop [0]{.\EOS\space}%
\providecommand \EOS [0]{\spacefactor3000\relax}%
\providecommand \BibitemShut  [1]{\csname bibitem#1\endcsname}%
\let\auto@bib@innerbib\@empty
\bibitem [{\citenamefont {Yip}(2005)}]{Yip2005}%
  \BibitemOpen
  \bibfield  {author} {\bibinfo {author} {\bibfnamefont {S.}~\bibnamefont {Yip}},\ }\bibinfo {title} {Introduction},\ in\ \href {https://doi.org/10.1007/978-1-4020-3286-8_1} {\emph {\bibinfo {booktitle} {Handbook of Materials Modeling: Methods}}},\ \bibinfo {editor} {edited by\ \bibinfo {editor} {\bibfnamefont {S.}~\bibnamefont {Yip}}}\ (\bibinfo  {publisher} {Springer Netherlands},\ \bibinfo {address} {Dordrecht},\ \bibinfo {year} {2005})\ pp.\ \bibinfo {pages} {1--5}\BibitemShut {NoStop}%
\bibitem [{\citenamefont {Carter}(2008)}]{carter2008challenges}%
  \BibitemOpen
  \bibfield  {author} {\bibinfo {author} {\bibfnamefont {E.~A.}\ \bibnamefont {Carter}},\ }\bibfield  {title} {\bibinfo {title} {Challenges in modeling materials properties without experimental input},\ }\href@noop {} {\bibfield  {journal} {\bibinfo  {journal} {Science}\ }\textbf {\bibinfo {volume} {321}},\ \bibinfo {pages} {800} (\bibinfo {year} {2008})}\BibitemShut {NoStop}%
\bibitem [{\citenamefont {Kulik}\ \emph {et~al.}(2022)\citenamefont {Kulik}, \citenamefont {Hammerschmidt}, \citenamefont {Schmidt}, \citenamefont {Botti}, \citenamefont {Marques}, \citenamefont {Boley}, \citenamefont {Scheffler}, \citenamefont {Todorovi{\'c}}, \citenamefont {Rinke}, \citenamefont {Oses} \emph {et~al.}}]{kulik2022roadmap}%
  \BibitemOpen
  \bibfield  {author} {\bibinfo {author} {\bibfnamefont {H.~J.}\ \bibnamefont {Kulik}}, \bibinfo {author} {\bibfnamefont {T.}~\bibnamefont {Hammerschmidt}}, \bibinfo {author} {\bibfnamefont {J.}~\bibnamefont {Schmidt}}, \bibinfo {author} {\bibfnamefont {S.}~\bibnamefont {Botti}}, \bibinfo {author} {\bibfnamefont {M.~A.}\ \bibnamefont {Marques}}, \bibinfo {author} {\bibfnamefont {M.}~\bibnamefont {Boley}}, \bibinfo {author} {\bibfnamefont {M.}~\bibnamefont {Scheffler}}, \bibinfo {author} {\bibfnamefont {M.}~\bibnamefont {Todorovi{\'c}}}, \bibinfo {author} {\bibfnamefont {P.}~\bibnamefont {Rinke}}, \bibinfo {author} {\bibfnamefont {C.}~\bibnamefont {Oses}}, \emph {et~al.},\ }\bibfield  {title} {\bibinfo {title} {Roadmap on machine learning in electronic structure},\ }\href@noop {} {\bibfield  {journal} {\bibinfo  {journal} {Electronic Structure}\ }\textbf {\bibinfo {volume} {4}},\ \bibinfo {pages} {023004} (\bibinfo {year} {2022})}\BibitemShut {NoStop}%
\bibitem [{\citenamefont {Dral}(2022)}]{dral2022quantum}%
  \BibitemOpen
  \bibfield  {author} {\bibinfo {author} {\bibfnamefont {P.~O.}\ \bibnamefont {Dral}},\ }\href@noop {} {\emph {\bibinfo {title} {Quantum Chemistry in the Age of Machine Learning}}}\ (\bibinfo  {publisher} {Elsevier},\ \bibinfo {year} {2022})\BibitemShut {NoStop}%
\bibitem [{\citenamefont {Takamoto}\ \emph {et~al.}(2022{\natexlab{a}})\citenamefont {Takamoto}, \citenamefont {Izumi},\ and\ \citenamefont {Li}}]{TakamotoIL22}%
  \BibitemOpen
  \bibfield  {author} {\bibinfo {author} {\bibfnamefont {S.}~\bibnamefont {Takamoto}}, \bibinfo {author} {\bibfnamefont {S.}~\bibnamefont {Izumi}},\ and\ \bibinfo {author} {\bibfnamefont {J.}~\bibnamefont {Li}},\ }\bibfield  {title} {\bibinfo {title} {Teanet: Universal neural network interatomic potential inspired by iterative electronic relaxations},\ }\href@noop {} {\bibfield  {journal} {\bibinfo  {journal} {Comput. Mater. Sci.}\ }\textbf {\bibinfo {volume} {207}},\ \bibinfo {pages} {111280} (\bibinfo {year} {2022}{\natexlab{a}})}\BibitemShut {NoStop}%
\bibitem [{\citenamefont {Takamoto}\ \emph {et~al.}(2023)\citenamefont {Takamoto}, \citenamefont {Okanohara}, \citenamefont {Li},\ and\ \citenamefont {Li}}]{TakamotoOLL23}%
  \BibitemOpen
  \bibfield  {author} {\bibinfo {author} {\bibfnamefont {S.}~\bibnamefont {Takamoto}}, \bibinfo {author} {\bibfnamefont {D.}~\bibnamefont {Okanohara}}, \bibinfo {author} {\bibfnamefont {Q.}~\bibnamefont {Li}},\ and\ \bibinfo {author} {\bibfnamefont {J.}~\bibnamefont {Li}},\ }\bibfield  {title} {\bibinfo {title} {Towards universal neural network interatomic potential},\ }\href@noop {} {\bibfield  {journal} {\bibinfo  {journal} {J. Materiomics}\ }\textbf {\bibinfo {volume} {9}},\ \bibinfo {pages} {447} (\bibinfo {year} {2023})}\BibitemShut {NoStop}%
\bibitem [{\citenamefont {Batzner}\ \emph {et~al.}(2022)\citenamefont {Batzner}, \citenamefont {Musaelian}, \citenamefont {Sun}, \citenamefont {Geiger}, \citenamefont {Mailoa}, \citenamefont {Kornbluth}, \citenamefont {Molinari}, \citenamefont {Smidt},\ and\ \citenamefont {Kozinsky}}]{batzner20223}%
  \BibitemOpen
  \bibfield  {author} {\bibinfo {author} {\bibfnamefont {S.}~\bibnamefont {Batzner}}, \bibinfo {author} {\bibfnamefont {A.}~\bibnamefont {Musaelian}}, \bibinfo {author} {\bibfnamefont {L.}~\bibnamefont {Sun}}, \bibinfo {author} {\bibfnamefont {M.}~\bibnamefont {Geiger}}, \bibinfo {author} {\bibfnamefont {J.~P.}\ \bibnamefont {Mailoa}}, \bibinfo {author} {\bibfnamefont {M.}~\bibnamefont {Kornbluth}}, \bibinfo {author} {\bibfnamefont {N.}~\bibnamefont {Molinari}}, \bibinfo {author} {\bibfnamefont {T.~E.}\ \bibnamefont {Smidt}},\ and\ \bibinfo {author} {\bibfnamefont {B.}~\bibnamefont {Kozinsky}},\ }\bibfield  {title} {\bibinfo {title} {E (3)-equivariant graph neural networks for data-efficient and accurate interatomic potentials},\ }\href@noop {} {\bibfield  {journal} {\bibinfo  {journal} {Nature communications}\ }\textbf {\bibinfo {volume} {13}},\ \bibinfo {pages} {2453} (\bibinfo {year} {2022})}\BibitemShut {NoStop}%
\bibitem [{\citenamefont {Zhang}\ \emph {et~al.}(2018{\natexlab{a}})\citenamefont {Zhang}, \citenamefont {Han}, \citenamefont {Wang}, \citenamefont {Car},\ and\ \citenamefont {Weinan}}]{zhang2018deep}%
  \BibitemOpen
  \bibfield  {author} {\bibinfo {author} {\bibfnamefont {L.}~\bibnamefont {Zhang}}, \bibinfo {author} {\bibfnamefont {J.}~\bibnamefont {Han}}, \bibinfo {author} {\bibfnamefont {H.}~\bibnamefont {Wang}}, \bibinfo {author} {\bibfnamefont {R.}~\bibnamefont {Car}},\ and\ \bibinfo {author} {\bibfnamefont {E.}~\bibnamefont {Weinan}},\ }\bibfield  {title} {\bibinfo {title} {Deep potential molecular dynamics: a scalable model with the accuracy of quantum mechanics},\ }\href@noop {} {\bibfield  {journal} {\bibinfo  {journal} {Physical review letters}\ }\textbf {\bibinfo {volume} {120}},\ \bibinfo {pages} {143001} (\bibinfo {year} {2018}{\natexlab{a}})}\BibitemShut {NoStop}%
\bibitem [{\citenamefont {Merchant}\ \emph {et~al.}(2023)\citenamefont {Merchant}, \citenamefont {Batzner}, \citenamefont {Schoenholz}, \citenamefont {Aykol}, \citenamefont {Cheon},\ and\ \citenamefont {Cubuk}}]{merchant2023scaling}%
  \BibitemOpen
  \bibfield  {author} {\bibinfo {author} {\bibfnamefont {A.}~\bibnamefont {Merchant}}, \bibinfo {author} {\bibfnamefont {S.}~\bibnamefont {Batzner}}, \bibinfo {author} {\bibfnamefont {S.~S.}\ \bibnamefont {Schoenholz}}, \bibinfo {author} {\bibfnamefont {M.}~\bibnamefont {Aykol}}, \bibinfo {author} {\bibfnamefont {G.}~\bibnamefont {Cheon}},\ and\ \bibinfo {author} {\bibfnamefont {E.~D.}\ \bibnamefont {Cubuk}},\ }\bibfield  {title} {\bibinfo {title} {Scaling deep learning for materials discovery},\ }\href@noop {} {\bibfield  {journal} {\bibinfo  {journal} {Nature}\ }\textbf {\bibinfo {volume} {624}},\ \bibinfo {pages} {80} (\bibinfo {year} {2023})}\BibitemShut {NoStop}%
\bibitem [{\citenamefont {Chen}\ and\ \citenamefont {Ong}(2022)}]{chen2022universal}%
  \BibitemOpen
  \bibfield  {author} {\bibinfo {author} {\bibfnamefont {C.}~\bibnamefont {Chen}}\ and\ \bibinfo {author} {\bibfnamefont {S.~P.}\ \bibnamefont {Ong}},\ }\bibfield  {title} {\bibinfo {title} {A universal graph deep learning interatomic potential for the periodic table},\ }\href@noop {} {\bibfield  {journal} {\bibinfo  {journal} {Nature Computational Science}\ }\textbf {\bibinfo {volume} {2}},\ \bibinfo {pages} {718} (\bibinfo {year} {2022})}\BibitemShut {NoStop}%
\bibitem [{\citenamefont {Takamoto}\ \emph {et~al.}(2022{\natexlab{b}})\citenamefont {Takamoto}, \citenamefont {Shinagawa}, \citenamefont {Motoki}, \citenamefont {Nakago}, \citenamefont {Li}, \citenamefont {Kurata}, \citenamefont {Watanabe}, \citenamefont {Yayama}, \citenamefont {Iriguchi}, \citenamefont {Asano} \emph {et~al.}}]{takamoto2022towards}%
  \BibitemOpen
  \bibfield  {author} {\bibinfo {author} {\bibfnamefont {S.}~\bibnamefont {Takamoto}}, \bibinfo {author} {\bibfnamefont {C.}~\bibnamefont {Shinagawa}}, \bibinfo {author} {\bibfnamefont {D.}~\bibnamefont {Motoki}}, \bibinfo {author} {\bibfnamefont {K.}~\bibnamefont {Nakago}}, \bibinfo {author} {\bibfnamefont {W.}~\bibnamefont {Li}}, \bibinfo {author} {\bibfnamefont {I.}~\bibnamefont {Kurata}}, \bibinfo {author} {\bibfnamefont {T.}~\bibnamefont {Watanabe}}, \bibinfo {author} {\bibfnamefont {Y.}~\bibnamefont {Yayama}}, \bibinfo {author} {\bibfnamefont {H.}~\bibnamefont {Iriguchi}}, \bibinfo {author} {\bibfnamefont {Y.}~\bibnamefont {Asano}}, \emph {et~al.},\ }\bibfield  {title} {\bibinfo {title} {Towards universal neural network potential for material discovery applicable to arbitrary combination of 45 elements},\ }\href@noop {} {\bibfield  {journal} {\bibinfo  {journal} {Nature Communications}\ }\textbf {\bibinfo {volume} {13}},\ \bibinfo {pages} {2991} (\bibinfo {year} {2022}{\natexlab{b}})}\BibitemShut
  {NoStop}%
\bibitem [{\citenamefont {Smith}\ \emph {et~al.}(2019)\citenamefont {Smith}, \citenamefont {Nebgen}, \citenamefont {Zubatyuk}, \citenamefont {Lubbers}, \citenamefont {Devereux}, \citenamefont {Barros}, \citenamefont {Tretiak}, \citenamefont {Isayev},\ and\ \citenamefont {Roitberg}}]{smith2019approaching}%
  \BibitemOpen
  \bibfield  {author} {\bibinfo {author} {\bibfnamefont {J.~S.}\ \bibnamefont {Smith}}, \bibinfo {author} {\bibfnamefont {B.~T.}\ \bibnamefont {Nebgen}}, \bibinfo {author} {\bibfnamefont {R.}~\bibnamefont {Zubatyuk}}, \bibinfo {author} {\bibfnamefont {N.}~\bibnamefont {Lubbers}}, \bibinfo {author} {\bibfnamefont {C.}~\bibnamefont {Devereux}}, \bibinfo {author} {\bibfnamefont {K.}~\bibnamefont {Barros}}, \bibinfo {author} {\bibfnamefont {S.}~\bibnamefont {Tretiak}}, \bibinfo {author} {\bibfnamefont {O.}~\bibnamefont {Isayev}},\ and\ \bibinfo {author} {\bibfnamefont {A.~E.}\ \bibnamefont {Roitberg}},\ }\bibfield  {title} {\bibinfo {title} {Approaching coupled cluster accuracy with a general-purpose neural network potential through transfer learning},\ }\href@noop {} {\bibfield  {journal} {\bibinfo  {journal} {Nature communications}\ }\textbf {\bibinfo {volume} {10}},\ \bibinfo {pages} {2903} (\bibinfo {year} {2019})}\BibitemShut {NoStop}%
\bibitem [{\citenamefont {Zheng}\ \emph {et~al.}(2021)\citenamefont {Zheng}, \citenamefont {Zubatyuk}, \citenamefont {Wu}, \citenamefont {Isayev},\ and\ \citenamefont {Dral}}]{zheng2021artificial}%
  \BibitemOpen
  \bibfield  {author} {\bibinfo {author} {\bibfnamefont {P.}~\bibnamefont {Zheng}}, \bibinfo {author} {\bibfnamefont {R.}~\bibnamefont {Zubatyuk}}, \bibinfo {author} {\bibfnamefont {W.}~\bibnamefont {Wu}}, \bibinfo {author} {\bibfnamefont {O.}~\bibnamefont {Isayev}},\ and\ \bibinfo {author} {\bibfnamefont {P.~O.}\ \bibnamefont {Dral}},\ }\bibfield  {title} {\bibinfo {title} {Artificial intelligence-enhanced quantum chemical method with broad applicability},\ }\href@noop {} {\bibfield  {journal} {\bibinfo  {journal} {Nature communications}\ }\textbf {\bibinfo {volume} {12}},\ \bibinfo {pages} {7022} (\bibinfo {year} {2021})}\BibitemShut {NoStop}%
\bibitem [{\citenamefont {Kirkpatrick}\ \emph {et~al.}(2021)\citenamefont {Kirkpatrick}, \citenamefont {McMorrow}, \citenamefont {Turban}, \citenamefont {Gaunt}, \citenamefont {Spencer}, \citenamefont {Matthews}, \citenamefont {Obika}, \citenamefont {Thiry}, \citenamefont {Fortunato}, \citenamefont {Pfau} \emph {et~al.}}]{kirkpatrick2021pushing}%
  \BibitemOpen
  \bibfield  {author} {\bibinfo {author} {\bibfnamefont {J.}~\bibnamefont {Kirkpatrick}}, \bibinfo {author} {\bibfnamefont {B.}~\bibnamefont {McMorrow}}, \bibinfo {author} {\bibfnamefont {D.~H.}\ \bibnamefont {Turban}}, \bibinfo {author} {\bibfnamefont {A.~L.}\ \bibnamefont {Gaunt}}, \bibinfo {author} {\bibfnamefont {J.~S.}\ \bibnamefont {Spencer}}, \bibinfo {author} {\bibfnamefont {A.~G.}\ \bibnamefont {Matthews}}, \bibinfo {author} {\bibfnamefont {A.}~\bibnamefont {Obika}}, \bibinfo {author} {\bibfnamefont {L.}~\bibnamefont {Thiry}}, \bibinfo {author} {\bibfnamefont {M.}~\bibnamefont {Fortunato}}, \bibinfo {author} {\bibfnamefont {D.}~\bibnamefont {Pfau}}, \emph {et~al.},\ }\bibfield  {title} {\bibinfo {title} {Pushing the frontiers of density functionals by solving the fractional electron problem},\ }\href@noop {} {\bibfield  {journal} {\bibinfo  {journal} {Science}\ }\textbf {\bibinfo {volume} {374}},\ \bibinfo {pages} {1385} (\bibinfo {year} {2021})}\BibitemShut {NoStop}%
\bibitem [{\citenamefont {Pederson}\ \emph {et~al.}(2022)\citenamefont {Pederson}, \citenamefont {Kalita},\ and\ \citenamefont {Burke}}]{pederson2022machine}%
  \BibitemOpen
  \bibfield  {author} {\bibinfo {author} {\bibfnamefont {R.}~\bibnamefont {Pederson}}, \bibinfo {author} {\bibfnamefont {B.}~\bibnamefont {Kalita}},\ and\ \bibinfo {author} {\bibfnamefont {K.}~\bibnamefont {Burke}},\ }\bibfield  {title} {\bibinfo {title} {Machine learning and density functional theory},\ }\href@noop {} {\bibfield  {journal} {\bibinfo  {journal} {Nature Reviews Physics}\ }\textbf {\bibinfo {volume} {4}},\ \bibinfo {pages} {357} (\bibinfo {year} {2022})}\BibitemShut {NoStop}%
\bibitem [{\citenamefont {Bogojeski}\ \emph {et~al.}(2020)\citenamefont {Bogojeski}, \citenamefont {Vogt-Maranto}, \citenamefont {Tuckerman}, \citenamefont {M{\"u}ller},\ and\ \citenamefont {Burke}}]{bogojeski2020quantum}%
  \BibitemOpen
  \bibfield  {author} {\bibinfo {author} {\bibfnamefont {M.}~\bibnamefont {Bogojeski}}, \bibinfo {author} {\bibfnamefont {L.}~\bibnamefont {Vogt-Maranto}}, \bibinfo {author} {\bibfnamefont {M.~E.}\ \bibnamefont {Tuckerman}}, \bibinfo {author} {\bibfnamefont {K.-R.}\ \bibnamefont {M{\"u}ller}},\ and\ \bibinfo {author} {\bibfnamefont {K.}~\bibnamefont {Burke}},\ }\bibfield  {title} {\bibinfo {title} {Quantum chemical accuracy from density functional approximations via machine learning},\ }\href@noop {} {\bibfield  {journal} {\bibinfo  {journal} {Nature communications}\ }\textbf {\bibinfo {volume} {11}},\ \bibinfo {pages} {5223} (\bibinfo {year} {2020})}\BibitemShut {NoStop}%
\bibitem [{\citenamefont {Bystrom}\ and\ \citenamefont {Kozinsky}(2022)}]{bystrom2022cider}%
  \BibitemOpen
  \bibfield  {author} {\bibinfo {author} {\bibfnamefont {K.}~\bibnamefont {Bystrom}}\ and\ \bibinfo {author} {\bibfnamefont {B.}~\bibnamefont {Kozinsky}},\ }\bibfield  {title} {\bibinfo {title} {Cider: An expressive, nonlocal feature set for machine learning density functionals with exact constraints},\ }\href@noop {} {\bibfield  {journal} {\bibinfo  {journal} {Journal of Chemical Theory and Computation}\ }\textbf {\bibinfo {volume} {18}},\ \bibinfo {pages} {2180} (\bibinfo {year} {2022})}\BibitemShut {NoStop}%
\bibitem [{\citenamefont {Helgaker}\ \emph {et~al.}(2013)\citenamefont {Helgaker}, \citenamefont {Jorgensen},\ and\ \citenamefont {Olsen}}]{helgaker2013molecular}%
  \BibitemOpen
  \bibfield  {author} {\bibinfo {author} {\bibfnamefont {T.}~\bibnamefont {Helgaker}}, \bibinfo {author} {\bibfnamefont {P.}~\bibnamefont {Jorgensen}},\ and\ \bibinfo {author} {\bibfnamefont {J.}~\bibnamefont {Olsen}},\ }\href@noop {} {\emph {\bibinfo {title} {Molecular electronic-structure theory}}}\ (\bibinfo  {publisher} {John Wiley \& Sons},\ \bibinfo {year} {2013})\BibitemShut {NoStop}%
\bibitem [{\citenamefont {Sch{\"u}tt}\ \emph {et~al.}(2019)\citenamefont {Sch{\"u}tt}, \citenamefont {Gastegger}, \citenamefont {Tkatchenko}, \citenamefont {M{\"u}ller},\ and\ \citenamefont {Maurer}}]{schutt2019unifying}%
  \BibitemOpen
  \bibfield  {author} {\bibinfo {author} {\bibfnamefont {K.~T.}\ \bibnamefont {Sch{\"u}tt}}, \bibinfo {author} {\bibfnamefont {M.}~\bibnamefont {Gastegger}}, \bibinfo {author} {\bibfnamefont {A.}~\bibnamefont {Tkatchenko}}, \bibinfo {author} {\bibfnamefont {K.-R.}\ \bibnamefont {M{\"u}ller}},\ and\ \bibinfo {author} {\bibfnamefont {R.~J.}\ \bibnamefont {Maurer}},\ }\bibfield  {title} {\bibinfo {title} {Unifying machine learning and quantum chemistry with a deep neural network for molecular wavefunctions},\ }\href@noop {} {\bibfield  {journal} {\bibinfo  {journal} {Nature communications}\ }\textbf {\bibinfo {volume} {10}},\ \bibinfo {pages} {5024} (\bibinfo {year} {2019})}\BibitemShut {NoStop}%
\bibitem [{\citenamefont {Shao}\ \emph {et~al.}(2023)\citenamefont {Shao}, \citenamefont {Paetow}, \citenamefont {Tuckerman},\ and\ \citenamefont {Pavanello}}]{shao2023machine}%
  \BibitemOpen
  \bibfield  {author} {\bibinfo {author} {\bibfnamefont {X.}~\bibnamefont {Shao}}, \bibinfo {author} {\bibfnamefont {L.}~\bibnamefont {Paetow}}, \bibinfo {author} {\bibfnamefont {M.~E.}\ \bibnamefont {Tuckerman}},\ and\ \bibinfo {author} {\bibfnamefont {M.}~\bibnamefont {Pavanello}},\ }\bibfield  {title} {\bibinfo {title} {Machine learning electronic structure methods based on the one-electron reduced density matrix},\ }\href@noop {} {\bibfield  {journal} {\bibinfo  {journal} {Nature communications}\ }\textbf {\bibinfo {volume} {14}},\ \bibinfo {pages} {6281} (\bibinfo {year} {2023})}\BibitemShut {NoStop}%
\bibitem [{\citenamefont {Feng}\ \emph {et~al.}(2023)\citenamefont {Feng}, \citenamefont {Xi}, \citenamefont {Zhang}, \citenamefont {Jiang},\ and\ \citenamefont {Zhou}}]{feng2023accurate}%
  \BibitemOpen
  \bibfield  {author} {\bibinfo {author} {\bibfnamefont {C.}~\bibnamefont {Feng}}, \bibinfo {author} {\bibfnamefont {J.}~\bibnamefont {Xi}}, \bibinfo {author} {\bibfnamefont {Y.}~\bibnamefont {Zhang}}, \bibinfo {author} {\bibfnamefont {B.}~\bibnamefont {Jiang}},\ and\ \bibinfo {author} {\bibfnamefont {Y.}~\bibnamefont {Zhou}},\ }\bibfield  {title} {\bibinfo {title} {Accurate and interpretable dipole interaction model-based machine learning for molecular polarizability},\ }\href@noop {} {\bibfield  {journal} {\bibinfo  {journal} {Journal of Chemical Theory and Computation}\ }\textbf {\bibinfo {volume} {19}},\ \bibinfo {pages} {1207} (\bibinfo {year} {2023})}\BibitemShut {NoStop}%
\bibitem [{\citenamefont {Fan}\ \emph {et~al.}(2022)\citenamefont {Fan}, \citenamefont {McSloy}, \citenamefont {Aradi}, \citenamefont {Yam},\ and\ \citenamefont {Frauenheim}}]{fan2022obtaining}%
  \BibitemOpen
  \bibfield  {author} {\bibinfo {author} {\bibfnamefont {G.}~\bibnamefont {Fan}}, \bibinfo {author} {\bibfnamefont {A.}~\bibnamefont {McSloy}}, \bibinfo {author} {\bibfnamefont {B.}~\bibnamefont {Aradi}}, \bibinfo {author} {\bibfnamefont {C.-Y.}\ \bibnamefont {Yam}},\ and\ \bibinfo {author} {\bibfnamefont {T.}~\bibnamefont {Frauenheim}},\ }\bibfield  {title} {\bibinfo {title} {Obtaining electronic properties of molecules through combining density functional tight binding with machine learning},\ }\href@noop {} {\bibfield  {journal} {\bibinfo  {journal} {The Journal of Physical Chemistry Letters}\ }\textbf {\bibinfo {volume} {13}},\ \bibinfo {pages} {10132} (\bibinfo {year} {2022})}\BibitemShut {NoStop}%
\bibitem [{\citenamefont {Cignoni}\ \emph {et~al.}(2023)\citenamefont {Cignoni}, \citenamefont {Suman}, \citenamefont {Nigam}, \citenamefont {Cupellini}, \citenamefont {Mennucci},\ and\ \citenamefont {Ceriotti}}]{cignoni2023electronic}%
  \BibitemOpen
  \bibfield  {author} {\bibinfo {author} {\bibfnamefont {E.}~\bibnamefont {Cignoni}}, \bibinfo {author} {\bibfnamefont {D.}~\bibnamefont {Suman}}, \bibinfo {author} {\bibfnamefont {J.}~\bibnamefont {Nigam}}, \bibinfo {author} {\bibfnamefont {L.}~\bibnamefont {Cupellini}}, \bibinfo {author} {\bibfnamefont {B.}~\bibnamefont {Mennucci}},\ and\ \bibinfo {author} {\bibfnamefont {M.}~\bibnamefont {Ceriotti}},\ }\bibfield  {title} {\bibinfo {title} {Electronic excited states from physically constrained machine learning},\ }\href@noop {} {\bibfield  {journal} {\bibinfo  {journal} {ACS Central Science}\ } (\bibinfo {year} {2023})}\BibitemShut {NoStop}%
\bibitem [{\citenamefont {Dral}\ and\ \citenamefont {Barbatti}(2021)}]{dral2021molecular}%
  \BibitemOpen
  \bibfield  {author} {\bibinfo {author} {\bibfnamefont {P.~O.}\ \bibnamefont {Dral}}\ and\ \bibinfo {author} {\bibfnamefont {M.}~\bibnamefont {Barbatti}},\ }\bibfield  {title} {\bibinfo {title} {Molecular excited states through a machine learning lens},\ }\href@noop {} {\bibfield  {journal} {\bibinfo  {journal} {Nature Reviews Chemistry}\ }\textbf {\bibinfo {volume} {5}},\ \bibinfo {pages} {388} (\bibinfo {year} {2021})}\BibitemShut {NoStop}%
\bibitem [{\citenamefont {Li}\ \emph {et~al.}(2022)\citenamefont {Li}, \citenamefont {Wang}, \citenamefont {Zou}, \citenamefont {Ye}, \citenamefont {Xu}, \citenamefont {Gong}, \citenamefont {Duan},\ and\ \citenamefont {Xu}}]{li2022deep}%
  \BibitemOpen
  \bibfield  {author} {\bibinfo {author} {\bibfnamefont {H.}~\bibnamefont {Li}}, \bibinfo {author} {\bibfnamefont {Z.}~\bibnamefont {Wang}}, \bibinfo {author} {\bibfnamefont {N.}~\bibnamefont {Zou}}, \bibinfo {author} {\bibfnamefont {M.}~\bibnamefont {Ye}}, \bibinfo {author} {\bibfnamefont {R.}~\bibnamefont {Xu}}, \bibinfo {author} {\bibfnamefont {X.}~\bibnamefont {Gong}}, \bibinfo {author} {\bibfnamefont {W.}~\bibnamefont {Duan}},\ and\ \bibinfo {author} {\bibfnamefont {Y.}~\bibnamefont {Xu}},\ }\bibfield  {title} {\bibinfo {title} {Deep-learning density functional theory hamiltonian for efficient ab initio electronic-structure calculation},\ }\href@noop {} {\bibfield  {journal} {\bibinfo  {journal} {Nature Computational Science}\ }\textbf {\bibinfo {volume} {2}},\ \bibinfo {pages} {367} (\bibinfo {year} {2022})}\BibitemShut {NoStop}%
\bibitem [{\citenamefont {Gong}\ \emph {et~al.}(2023)\citenamefont {Gong}, \citenamefont {Li}, \citenamefont {Zou}, \citenamefont {Xu}, \citenamefont {Duan},\ and\ \citenamefont {Xu}}]{gong2023general}%
  \BibitemOpen
  \bibfield  {author} {\bibinfo {author} {\bibfnamefont {X.}~\bibnamefont {Gong}}, \bibinfo {author} {\bibfnamefont {H.}~\bibnamefont {Li}}, \bibinfo {author} {\bibfnamefont {N.}~\bibnamefont {Zou}}, \bibinfo {author} {\bibfnamefont {R.}~\bibnamefont {Xu}}, \bibinfo {author} {\bibfnamefont {W.}~\bibnamefont {Duan}},\ and\ \bibinfo {author} {\bibfnamefont {Y.}~\bibnamefont {Xu}},\ }\bibfield  {title} {\bibinfo {title} {General framework for e (3)-equivariant neural network representation of density functional theory hamiltonian},\ }\href@noop {} {\bibfield  {journal} {\bibinfo  {journal} {Nature Communications}\ }\textbf {\bibinfo {volume} {14}},\ \bibinfo {pages} {2848} (\bibinfo {year} {2023})}\BibitemShut {NoStop}%
\bibitem [{\citenamefont {Unke}\ \emph {et~al.}(2021)\citenamefont {Unke}, \citenamefont {Bogojeski}, \citenamefont {Gastegger}, \citenamefont {Geiger}, \citenamefont {Smidt},\ and\ \citenamefont {M{\"u}ller}}]{unke2021se}%
  \BibitemOpen
  \bibfield  {author} {\bibinfo {author} {\bibfnamefont {O.}~\bibnamefont {Unke}}, \bibinfo {author} {\bibfnamefont {M.}~\bibnamefont {Bogojeski}}, \bibinfo {author} {\bibfnamefont {M.}~\bibnamefont {Gastegger}}, \bibinfo {author} {\bibfnamefont {M.}~\bibnamefont {Geiger}}, \bibinfo {author} {\bibfnamefont {T.}~\bibnamefont {Smidt}},\ and\ \bibinfo {author} {\bibfnamefont {K.-R.}\ \bibnamefont {M{\"u}ller}},\ }\bibfield  {title} {\bibinfo {title} {Se (3)-equivariant prediction of molecular wavefunctions and electronic densities},\ }\href@noop {} {\bibfield  {journal} {\bibinfo  {journal} {Advances in Neural Information Processing Systems}\ }\textbf {\bibinfo {volume} {34}},\ \bibinfo {pages} {14434} (\bibinfo {year} {2021})}\BibitemShut {NoStop}%
\bibitem [{\citenamefont {Mardirossian}\ and\ \citenamefont {Head-Gordon}(2017)}]{mardirossian2017thirty}%
  \BibitemOpen
  \bibfield  {author} {\bibinfo {author} {\bibfnamefont {N.}~\bibnamefont {Mardirossian}}\ and\ \bibinfo {author} {\bibfnamefont {M.}~\bibnamefont {Head-Gordon}},\ }\bibfield  {title} {\bibinfo {title} {Thirty years of density functional theory in computational chemistry: an overview and extensive assessment of 200 density functionals},\ }\href@noop {} {\bibfield  {journal} {\bibinfo  {journal} {Molecular physics}\ }\textbf {\bibinfo {volume} {115}},\ \bibinfo {pages} {2315} (\bibinfo {year} {2017})}\BibitemShut {NoStop}%
\bibitem [{\citenamefont {Bartlett}\ and\ \citenamefont {Musia{\l}}(2007)}]{bartlett2007coupled}%
  \BibitemOpen
  \bibfield  {author} {\bibinfo {author} {\bibfnamefont {R.~J.}\ \bibnamefont {Bartlett}}\ and\ \bibinfo {author} {\bibfnamefont {M.}~\bibnamefont {Musia{\l}}},\ }\bibfield  {title} {\bibinfo {title} {Coupled-cluster theory in quantum chemistry},\ }\href@noop {} {\bibfield  {journal} {\bibinfo  {journal} {Reviews of Modern Physics}\ }\textbf {\bibinfo {volume} {79}},\ \bibinfo {pages} {291} (\bibinfo {year} {2007})}\BibitemShut {NoStop}%
\bibitem [{\citenamefont {Geiger}\ and\ \citenamefont {Smidt}(2022)}]{geiger2022e3nn}%
  \BibitemOpen
  \bibfield  {author} {\bibinfo {author} {\bibfnamefont {M.}~\bibnamefont {Geiger}}\ and\ \bibinfo {author} {\bibfnamefont {T.}~\bibnamefont {Smidt}},\ }\bibfield  {title} {\bibinfo {title} {e3nn: Euclidean neural networks},\ }\href@noop {} {\bibfield  {journal} {\bibinfo  {journal} {arXiv preprint arXiv:2207.09453}\ } (\bibinfo {year} {2022})}\BibitemShut {NoStop}%
\bibitem [{\citenamefont {Tirado-Rives}\ and\ \citenamefont {Jorgensen}(2008)}]{tirado2008performance}%
  \BibitemOpen
  \bibfield  {author} {\bibinfo {author} {\bibfnamefont {J.}~\bibnamefont {Tirado-Rives}}\ and\ \bibinfo {author} {\bibfnamefont {W.~L.}\ \bibnamefont {Jorgensen}},\ }\bibfield  {title} {\bibinfo {title} {Performance of b3lyp density functional methods for a large set of organic molecules},\ }\href@noop {} {\bibfield  {journal} {\bibinfo  {journal} {Journal of chemical theory and computation}\ }\textbf {\bibinfo {volume} {4}},\ \bibinfo {pages} {297} (\bibinfo {year} {2008})}\BibitemShut {NoStop}%
\bibitem [{\citenamefont {Dunning~Jr}(1989)}]{dunning1989gaussian}%
  \BibitemOpen
  \bibfield  {author} {\bibinfo {author} {\bibfnamefont {T.~H.}\ \bibnamefont {Dunning~Jr}},\ }\bibfield  {title} {\bibinfo {title} {Gaussian basis sets for use in correlated molecular calculations. i. the atoms boron through neon and hydrogen},\ }\href@noop {} {\bibfield  {journal} {\bibinfo  {journal} {The Journal of chemical physics}\ }\textbf {\bibinfo {volume} {90}},\ \bibinfo {pages} {1007} (\bibinfo {year} {1989})}\BibitemShut {NoStop}%
\bibitem [{\citenamefont {Neese}\ \emph {et~al.}(2020)\citenamefont {Neese}, \citenamefont {Wennmohs}, \citenamefont {Becker},\ and\ \citenamefont {Riplinger}}]{neese2020orca}%
  \BibitemOpen
  \bibfield  {author} {\bibinfo {author} {\bibfnamefont {F.}~\bibnamefont {Neese}}, \bibinfo {author} {\bibfnamefont {F.}~\bibnamefont {Wennmohs}}, \bibinfo {author} {\bibfnamefont {U.}~\bibnamefont {Becker}},\ and\ \bibinfo {author} {\bibfnamefont {C.}~\bibnamefont {Riplinger}},\ }\bibfield  {title} {\bibinfo {title} {The orca quantum chemistry program package},\ }\href@noop {} {\bibfield  {journal} {\bibinfo  {journal} {The Journal of chemical physics}\ }\textbf {\bibinfo {volume} {152}} (\bibinfo {year} {2020})}\BibitemShut {NoStop}%
\bibitem [{\citenamefont {Perdew}(1986)}]{perdew1986density}%
  \BibitemOpen
  \bibfield  {author} {\bibinfo {author} {\bibfnamefont {J.~P.}\ \bibnamefont {Perdew}},\ }\bibfield  {title} {\bibinfo {title} {Density-functional approximation for the correlation energy of the inhomogeneous electron gas},\ }\href@noop {} {\bibfield  {journal} {\bibinfo  {journal} {Physical review B}\ }\textbf {\bibinfo {volume} {33}},\ \bibinfo {pages} {8822} (\bibinfo {year} {1986})}\BibitemShut {NoStop}%
\bibitem [{\citenamefont {L{\"o}wdin}(1950)}]{lowdin1950non}%
  \BibitemOpen
  \bibfield  {author} {\bibinfo {author} {\bibfnamefont {P.-O.}\ \bibnamefont {L{\"o}wdin}},\ }\bibfield  {title} {\bibinfo {title} {On the non-orthogonality problem connected with the use of atomic wave functions in the theory of molecules and crystals},\ }\href@noop {} {\bibfield  {journal} {\bibinfo  {journal} {The Journal of Chemical Physics}\ }\textbf {\bibinfo {volume} {18}},\ \bibinfo {pages} {365} (\bibinfo {year} {1950})}\BibitemShut {NoStop}%
\bibitem [{\citenamefont {Mulliken}(1955)}]{mulliken1955electronic}%
  \BibitemOpen
  \bibfield  {author} {\bibinfo {author} {\bibfnamefont {R.~S.}\ \bibnamefont {Mulliken}},\ }\bibfield  {title} {\bibinfo {title} {Electronic population analysis on lcao--mo molecular wave functions. i},\ }\href@noop {} {\bibfield  {journal} {\bibinfo  {journal} {The Journal of chemical physics}\ }\textbf {\bibinfo {volume} {23}},\ \bibinfo {pages} {1833} (\bibinfo {year} {1955})}\BibitemShut {NoStop}%
\bibitem [{\citenamefont {Mayer}(2007)}]{mayer2007bond}%
  \BibitemOpen
  \bibfield  {author} {\bibinfo {author} {\bibfnamefont {I.}~\bibnamefont {Mayer}},\ }\bibfield  {title} {\bibinfo {title} {Bond order and valence indices: A personal account},\ }\href@noop {} {\bibfield  {journal} {\bibinfo  {journal} {Journal of computational chemistry}\ }\textbf {\bibinfo {volume} {28}},\ \bibinfo {pages} {204} (\bibinfo {year} {2007})}\BibitemShut {NoStop}%
\bibitem [{\citenamefont {Reuther}\ \emph {et~al.}(2018)\citenamefont {Reuther}, \citenamefont {Kepner}, \citenamefont {Byun}, \citenamefont {Samsi}, \citenamefont {Arcand}, \citenamefont {Bestor}, \citenamefont {Bergeron}, \citenamefont {Gadepally}, \citenamefont {Houle}, \citenamefont {Hubbell}, \citenamefont {Jones}, \citenamefont {Klein}, \citenamefont {Milechin}, \citenamefont {Mullen}, \citenamefont {Prout}, \citenamefont {Rosa}, \citenamefont {Yee},\ and\ \citenamefont {Michaleas}}]{reuther2018interactive}%
  \BibitemOpen
  \bibfield  {author} {\bibinfo {author} {\bibfnamefont {A.}~\bibnamefont {Reuther}}, \bibinfo {author} {\bibfnamefont {J.}~\bibnamefont {Kepner}}, \bibinfo {author} {\bibfnamefont {C.}~\bibnamefont {Byun}}, \bibinfo {author} {\bibfnamefont {S.}~\bibnamefont {Samsi}}, \bibinfo {author} {\bibfnamefont {W.}~\bibnamefont {Arcand}}, \bibinfo {author} {\bibfnamefont {D.}~\bibnamefont {Bestor}}, \bibinfo {author} {\bibfnamefont {B.}~\bibnamefont {Bergeron}}, \bibinfo {author} {\bibfnamefont {V.}~\bibnamefont {Gadepally}}, \bibinfo {author} {\bibfnamefont {M.}~\bibnamefont {Houle}}, \bibinfo {author} {\bibfnamefont {M.}~\bibnamefont {Hubbell}}, \bibinfo {author} {\bibfnamefont {M.}~\bibnamefont {Jones}}, \bibinfo {author} {\bibfnamefont {A.}~\bibnamefont {Klein}}, \bibinfo {author} {\bibfnamefont {L.}~\bibnamefont {Milechin}}, \bibinfo {author} {\bibfnamefont {J.}~\bibnamefont {Mullen}}, \bibinfo {author} {\bibfnamefont {A.}~\bibnamefont {Prout}}, \bibinfo {author} {\bibfnamefont {A.}~\bibnamefont {Rosa}}, \bibinfo
  {author} {\bibfnamefont {C.}~\bibnamefont {Yee}},\ and\ \bibinfo {author} {\bibfnamefont {P.}~\bibnamefont {Michaleas}},\ }\bibfield  {title} {\bibinfo {title} {Interactive supercomputing on 40,000 cores for machine learning and data analysis},\ }in\ \href@noop {} {\emph {\bibinfo {booktitle} {2018 IEEE High Performance extreme Computing Conference (HPEC)}}}\ (\bibinfo {organization} {IEEE},\ \bibinfo {year} {2018})\ pp.\ \bibinfo {pages} {1--6}\BibitemShut {NoStop}%
\bibitem [{\citenamefont {Kozuch}\ and\ \citenamefont {Martin}(2013)}]{kozuch2013spin}%
  \BibitemOpen
  \bibfield  {author} {\bibinfo {author} {\bibfnamefont {S.}~\bibnamefont {Kozuch}}\ and\ \bibinfo {author} {\bibfnamefont {J.~M.}\ \bibnamefont {Martin}},\ }\bibfield  {title} {\bibinfo {title} {Spin-component-scaled double hybrids: an extensive search for the best fifth-rung functionals blending dft and perturbation theory},\ }\href@noop {} {\bibfield  {journal} {\bibinfo  {journal} {Journal of Computational Chemistry}\ }\textbf {\bibinfo {volume} {34}},\ \bibinfo {pages} {2327} (\bibinfo {year} {2013})}\BibitemShut {NoStop}%
\bibitem [{\citenamefont {Dral}\ \emph {et~al.}(2024)\citenamefont {Dral}, \citenamefont {Ge}, \citenamefont {Hou}, \citenamefont {Zheng}, \citenamefont {Chen}, \citenamefont {Barbatti}, \citenamefont {Isayev}, \citenamefont {Wang}, \citenamefont {Xue}, \citenamefont {Pinheiro~Jr} \emph {et~al.}}]{dral2024mlatom}%
  \BibitemOpen
  \bibfield  {author} {\bibinfo {author} {\bibfnamefont {P.~O.}\ \bibnamefont {Dral}}, \bibinfo {author} {\bibfnamefont {F.}~\bibnamefont {Ge}}, \bibinfo {author} {\bibfnamefont {Y.-F.}\ \bibnamefont {Hou}}, \bibinfo {author} {\bibfnamefont {P.}~\bibnamefont {Zheng}}, \bibinfo {author} {\bibfnamefont {Y.}~\bibnamefont {Chen}}, \bibinfo {author} {\bibfnamefont {M.}~\bibnamefont {Barbatti}}, \bibinfo {author} {\bibfnamefont {O.}~\bibnamefont {Isayev}}, \bibinfo {author} {\bibfnamefont {C.}~\bibnamefont {Wang}}, \bibinfo {author} {\bibfnamefont {B.-X.}\ \bibnamefont {Xue}}, \bibinfo {author} {\bibfnamefont {M.}~\bibnamefont {Pinheiro~Jr}}, \emph {et~al.},\ }\bibfield  {title} {\bibinfo {title} {Mlatom 3: A platform for machine learning-enhanced computational chemistry simulations and workflows},\ }\href@noop {} {\bibfield  {journal} {\bibinfo  {journal} {Journal of Chemical Theory and Computation}\ } (\bibinfo {year} {2024})}\BibitemShut {NoStop}%
\bibitem [{\citenamefont {https://pytorch.org/docs/stable/generated/torch.linalg. eigh.html}(2023)}]{eigh}%
  \BibitemOpen
  \bibfield  {author} {\bibinfo {author} {\bibnamefont {https://pytorch.org/docs/stable/generated/torch.linalg. eigh.html}},\ }\href@noop {} {} (\bibinfo {year} {2023})\BibitemShut {NoStop}%
\bibitem [{\citenamefont {Kim}\ \emph {et~al.}(2022)\citenamefont {Kim}, \citenamefont {Chen}, \citenamefont {Cheng}, \citenamefont {Gindulyte}, \citenamefont {He}, \citenamefont {He}, \citenamefont {Li}, \citenamefont {Shoemaker}, \citenamefont {Thiessen}, \citenamefont {Yu}, \citenamefont {Zaslavsky}, \citenamefont {Zhang},\ and\ \citenamefont {Bolton}}]{pubchem}%
  \BibitemOpen
  \bibfield  {author} {\bibinfo {author} {\bibfnamefont {S.}~\bibnamefont {Kim}}, \bibinfo {author} {\bibfnamefont {J.}~\bibnamefont {Chen}}, \bibinfo {author} {\bibfnamefont {T.}~\bibnamefont {Cheng}}, \bibinfo {author} {\bibfnamefont {A.}~\bibnamefont {Gindulyte}}, \bibinfo {author} {\bibfnamefont {J.}~\bibnamefont {He}}, \bibinfo {author} {\bibfnamefont {S.}~\bibnamefont {He}}, \bibinfo {author} {\bibfnamefont {Q.}~\bibnamefont {Li}}, \bibinfo {author} {\bibfnamefont {B.~A.}\ \bibnamefont {Shoemaker}}, \bibinfo {author} {\bibfnamefont {P.~A.}\ \bibnamefont {Thiessen}}, \bibinfo {author} {\bibfnamefont {B.}~\bibnamefont {Yu}}, \bibinfo {author} {\bibfnamefont {L.}~\bibnamefont {Zaslavsky}}, \bibinfo {author} {\bibfnamefont {J.}~\bibnamefont {Zhang}},\ and\ \bibinfo {author} {\bibfnamefont {E.~E.}\ \bibnamefont {Bolton}},\ }\bibfield  {title} {\bibinfo {title} {{PubChem 2023 update}},\ }\href {https://doi.org/10.1093/nar/gkac956} {\bibfield  {journal} {\bibinfo  {journal} {Nucleic Acids Research}\ }\textbf
  {\bibinfo {volume} {51}},\ \bibinfo {pages} {D1373} (\bibinfo {year} {2022})},\ \Eprint {https://arxiv.org/abs/https://academic.oup.com/nar/article-pdf/51/D1/D1373/48441598/gkac956.pdf} {https://academic.oup.com/nar/article-pdf/51/D1/D1373/48441598/gkac956.pdf} \BibitemShut {NoStop}%
\bibitem [{\citenamefont {Krylov}(2008)}]{krylov2008equation}%
  \BibitemOpen
  \bibfield  {author} {\bibinfo {author} {\bibfnamefont {A.~I.}\ \bibnamefont {Krylov}},\ }\bibfield  {title} {\bibinfo {title} {Equation-of-motion coupled-cluster methods for open-shell and electronically excited species: The hitchhiker's guide to fock space},\ }\href@noop {} {\bibfield  {journal} {\bibinfo  {journal} {Annu. Rev. Phys. Chem.}\ }\textbf {\bibinfo {volume} {59}},\ \bibinfo {pages} {433} (\bibinfo {year} {2008})}\BibitemShut {NoStop}%
\bibitem [{\citenamefont {Caruana}(1997)}]{caruana1997multitask}%
  \BibitemOpen
  \bibfield  {author} {\bibinfo {author} {\bibfnamefont {R.}~\bibnamefont {Caruana}},\ }\bibfield  {title} {\bibinfo {title} {Multitask learning},\ }\href@noop {} {\bibfield  {journal} {\bibinfo  {journal} {Machine learning}\ }\textbf {\bibinfo {volume} {28}},\ \bibinfo {pages} {41} (\bibinfo {year} {1997})}\BibitemShut {NoStop}%
\bibitem [{\citenamefont {Slayden}\ and\ \citenamefont {Liebman}(2001)}]{slayden2001energetics}%
  \BibitemOpen
  \bibfield  {author} {\bibinfo {author} {\bibfnamefont {S.~W.}\ \bibnamefont {Slayden}}\ and\ \bibinfo {author} {\bibfnamefont {J.~F.}\ \bibnamefont {Liebman}},\ }\bibfield  {title} {\bibinfo {title} {The energetics of aromatic hydrocarbons: an experimental thermochemical perspective},\ }\href@noop {} {\bibfield  {journal} {\bibinfo  {journal} {Chemical reviews}\ }\textbf {\bibinfo {volume} {101}},\ \bibinfo {pages} {1541} (\bibinfo {year} {2001})}\BibitemShut {NoStop}%
\bibitem [{\citenamefont {Linstrom}\ and\ \citenamefont {W.G.~Mallard}(2024)}]{NISTwebbook}%
  \BibitemOpen
  \bibfield  {author} {\bibinfo {author} {\bibfnamefont {P.}~\bibnamefont {Linstrom}}\ and\ \bibinfo {author} {\bibfnamefont {E.}~\bibnamefont {W.G.~Mallard}},\ }\bibinfo {title} {Nist chemistry webbook, nist standard reference database number 69}\ (\bibinfo  {publisher} {National Institute of Standards and Technology, Gaithersburg MD, 20899},\ \bibinfo {year} {2024})\ Chap.\ \bibinfo {chapter} {Infrared Spectra}\BibitemShut {NoStop}%
\bibitem [{\citenamefont {Wilson~Jr}(1934)}]{wilson1934normal}%
  \BibitemOpen
  \bibfield  {author} {\bibinfo {author} {\bibfnamefont {E.~B.}\ \bibnamefont {Wilson~Jr}},\ }\bibfield  {title} {\bibinfo {title} {The normal modes and frequencies of vibration of the regular plane hexagon model of the benzene molecule},\ }\href@noop {} {\bibfield  {journal} {\bibinfo  {journal} {Physical Review}\ }\textbf {\bibinfo {volume} {45}},\ \bibinfo {pages} {706} (\bibinfo {year} {1934})}\BibitemShut {NoStop}%
\bibitem [{\citenamefont {Grem}\ \emph {et~al.}(1992)\citenamefont {Grem}, \citenamefont {Leditzky}, \citenamefont {Ullrich},\ and\ \citenamefont {Leising}}]{grem1992realization}%
  \BibitemOpen
  \bibfield  {author} {\bibinfo {author} {\bibfnamefont {G.}~\bibnamefont {Grem}}, \bibinfo {author} {\bibfnamefont {G.}~\bibnamefont {Leditzky}}, \bibinfo {author} {\bibfnamefont {B.}~\bibnamefont {Ullrich}},\ and\ \bibinfo {author} {\bibfnamefont {G.}~\bibnamefont {Leising}},\ }\bibfield  {title} {\bibinfo {title} {Realization of a blue-light-emitting device using poly (p-phenylene)},\ }\href@noop {} {\bibfield  {journal} {\bibinfo  {journal} {Advanced Materials}\ }\textbf {\bibinfo {volume} {4}},\ \bibinfo {pages} {36} (\bibinfo {year} {1992})}\BibitemShut {NoStop}%
\bibitem [{\citenamefont {Heeger}(2001)}]{heeger2001nobel}%
  \BibitemOpen
  \bibfield  {author} {\bibinfo {author} {\bibfnamefont {A.~J.}\ \bibnamefont {Heeger}},\ }\bibfield  {title} {\bibinfo {title} {Nobel lecture: Semiconducting and metallic polymers: The fourth generation of polymeric materials},\ }\href@noop {} {\bibfield  {journal} {\bibinfo  {journal} {Reviews of Modern Physics}\ }\textbf {\bibinfo {volume} {73}},\ \bibinfo {pages} {681} (\bibinfo {year} {2001})}\BibitemShut {NoStop}%
\bibitem [{\citenamefont {Otto}\ \emph {et~al.}(2004)\citenamefont {Otto}, \citenamefont {Piris}, \citenamefont {Martinez},\ and\ \citenamefont {Ladik}}]{otto2004dynamic}%
  \BibitemOpen
  \bibfield  {author} {\bibinfo {author} {\bibfnamefont {P.}~\bibnamefont {Otto}}, \bibinfo {author} {\bibfnamefont {M.}~\bibnamefont {Piris}}, \bibinfo {author} {\bibfnamefont {A.}~\bibnamefont {Martinez}},\ and\ \bibinfo {author} {\bibfnamefont {J.}~\bibnamefont {Ladik}},\ }\bibfield  {title} {\bibinfo {title} {Dynamic (hyper) polarizability calculations for polymers with linear and cyclic $\pi$-conjugated elementary cells},\ }\href@noop {} {\bibfield  {journal} {\bibinfo  {journal} {Synthetic metals}\ }\textbf {\bibinfo {volume} {141}},\ \bibinfo {pages} {277} (\bibinfo {year} {2004})}\BibitemShut {NoStop}%
\bibitem [{\citenamefont {Champagne}\ \emph {et~al.}(1998)\citenamefont {Champagne}, \citenamefont {Perpete}, \citenamefont {Van~Gisbergen}, \citenamefont {Baerends}, \citenamefont {Snijders}, \citenamefont {Soubra-Ghaoui}, \citenamefont {Robins},\ and\ \citenamefont {Kirtman}}]{champagne1998assessment}%
  \BibitemOpen
  \bibfield  {author} {\bibinfo {author} {\bibfnamefont {B.}~\bibnamefont {Champagne}}, \bibinfo {author} {\bibfnamefont {E.~A.}\ \bibnamefont {Perpete}}, \bibinfo {author} {\bibfnamefont {S.~J.}\ \bibnamefont {Van~Gisbergen}}, \bibinfo {author} {\bibfnamefont {E.-J.}\ \bibnamefont {Baerends}}, \bibinfo {author} {\bibfnamefont {J.~G.}\ \bibnamefont {Snijders}}, \bibinfo {author} {\bibfnamefont {C.}~\bibnamefont {Soubra-Ghaoui}}, \bibinfo {author} {\bibfnamefont {K.~A.}\ \bibnamefont {Robins}},\ and\ \bibinfo {author} {\bibfnamefont {B.}~\bibnamefont {Kirtman}},\ }\bibfield  {title} {\bibinfo {title} {Assessment of conventional density functional schemes for computing the polarizabilities and hyperpolarizabilities of conjugated oligomers: An ab initio investigation of polyacetylene chains},\ }\href@noop {} {\bibfield  {journal} {\bibinfo  {journal} {The Journal of chemical physics}\ }\textbf {\bibinfo {volume} {109}},\ \bibinfo {pages} {10489} (\bibinfo {year} {1998})}\BibitemShut {NoStop}%
\bibitem [{\citenamefont {Geffroy}\ \emph {et~al.}(2006)\citenamefont {Geffroy}, \citenamefont {Le~Roy},\ and\ \citenamefont {Prat}}]{geffroy2006organic}%
  \BibitemOpen
  \bibfield  {author} {\bibinfo {author} {\bibfnamefont {B.}~\bibnamefont {Geffroy}}, \bibinfo {author} {\bibfnamefont {P.}~\bibnamefont {Le~Roy}},\ and\ \bibinfo {author} {\bibfnamefont {C.}~\bibnamefont {Prat}},\ }\bibfield  {title} {\bibinfo {title} {Organic light-emitting diode (oled) technology: materials, devices and display technologies},\ }\href@noop {} {\bibfield  {journal} {\bibinfo  {journal} {Polymer international}\ }\textbf {\bibinfo {volume} {55}},\ \bibinfo {pages} {572} (\bibinfo {year} {2006})}\BibitemShut {NoStop}%
\bibitem [{\citenamefont {Zhang}\ \emph {et~al.}(2018{\natexlab{b}})\citenamefont {Zhang}, \citenamefont {Han}, \citenamefont {Wang}, \citenamefont {Saidi}, \citenamefont {Car} \emph {et~al.}}]{zhang2018end}%
  \BibitemOpen
  \bibfield  {author} {\bibinfo {author} {\bibfnamefont {L.}~\bibnamefont {Zhang}}, \bibinfo {author} {\bibfnamefont {J.}~\bibnamefont {Han}}, \bibinfo {author} {\bibfnamefont {H.}~\bibnamefont {Wang}}, \bibinfo {author} {\bibfnamefont {W.}~\bibnamefont {Saidi}}, \bibinfo {author} {\bibfnamefont {R.}~\bibnamefont {Car}}, \emph {et~al.},\ }\bibfield  {title} {\bibinfo {title} {End-to-end symmetry preserving inter-atomic potential energy model for finite and extended systems},\ }\href@noop {} {\bibfield  {journal} {\bibinfo  {journal} {Advances in neural information processing systems}\ }\textbf {\bibinfo {volume} {31}} (\bibinfo {year} {2018}{\natexlab{b}})}\BibitemShut {NoStop}%
\bibitem [{\citenamefont {Hanwell}\ \emph {et~al.}(2012)\citenamefont {Hanwell}, \citenamefont {Curtis}, \citenamefont {Lonie}, \citenamefont {Vandermeersch}, \citenamefont {Zurek},\ and\ \citenamefont {Hutchison}}]{hanwell2012avogadro}%
  \BibitemOpen
  \bibfield  {author} {\bibinfo {author} {\bibfnamefont {M.~D.}\ \bibnamefont {Hanwell}}, \bibinfo {author} {\bibfnamefont {D.~E.}\ \bibnamefont {Curtis}}, \bibinfo {author} {\bibfnamefont {D.~C.}\ \bibnamefont {Lonie}}, \bibinfo {author} {\bibfnamefont {T.}~\bibnamefont {Vandermeersch}}, \bibinfo {author} {\bibfnamefont {E.}~\bibnamefont {Zurek}},\ and\ \bibinfo {author} {\bibfnamefont {G.~R.}\ \bibnamefont {Hutchison}},\ }\bibfield  {title} {\bibinfo {title} {Avogadro: an advanced semantic chemical editor, visualization, and analysis platform},\ }\href@noop {} {\bibfield  {journal} {\bibinfo  {journal} {Journal of cheminformatics}\ }\textbf {\bibinfo {volume} {4}},\ \bibinfo {pages} {1} (\bibinfo {year} {2012})}\BibitemShut {NoStop}%
\bibitem [{\citenamefont {Kesharwani}\ \emph {et~al.}(2015)\citenamefont {Kesharwani}, \citenamefont {Brauer},\ and\ \citenamefont {Martin}}]{kesharwani2015frequency}%
  \BibitemOpen
  \bibfield  {author} {\bibinfo {author} {\bibfnamefont {M.~K.}\ \bibnamefont {Kesharwani}}, \bibinfo {author} {\bibfnamefont {B.}~\bibnamefont {Brauer}},\ and\ \bibinfo {author} {\bibfnamefont {J.~M.}\ \bibnamefont {Martin}},\ }\bibfield  {title} {\bibinfo {title} {Frequency and zero-point vibrational energy scale factors for double-hybrid density functionals (and other selected methods): can anharmonic force fields be avoided?},\ }\href@noop {} {\bibfield  {journal} {\bibinfo  {journal} {The Journal of Physical Chemistry A}\ }\textbf {\bibinfo {volume} {119}},\ \bibinfo {pages} {1701} (\bibinfo {year} {2015})}\BibitemShut {NoStop}%
\bibitem [{\citenamefont {Sun}\ \emph {et~al.}(2020)\citenamefont {Sun}, \citenamefont {Zhang}, \citenamefont {Banerjee}, \citenamefont {Bao}, \citenamefont {Barbry}, \citenamefont {Blunt}, \citenamefont {Bogdanov}, \citenamefont {Booth}, \citenamefont {Chen}, \citenamefont {Cui}, \citenamefont {Eriksen}, \citenamefont {Gao}, \citenamefont {Guo}, \citenamefont {Hermann}, \citenamefont {Hermes}, \citenamefont {Koh}, \citenamefont {Koval}, \citenamefont {Lehtola}, \citenamefont {Li}, \citenamefont {Liu}, \citenamefont {Mardirossian}, \citenamefont {McClain}, \citenamefont {Motta}, \citenamefont {Mussard}, \citenamefont {Pham}, \citenamefont {Pulkin}, \citenamefont {Purwanto}, \citenamefont {Robinson}, \citenamefont {Ronca}, \citenamefont {Sayfutyarova}, \citenamefont {Scheurer}, \citenamefont {Schurkus}, \citenamefont {Smith}, \citenamefont {Sun}, \citenamefont {Sun}, \citenamefont {Upadhyay}, \citenamefont {Wagner}, \citenamefont {Wang}, \citenamefont {White}, \citenamefont {Whitfield}, \citenamefont
  {Williamson}, \citenamefont {Wouters}, \citenamefont {Yang}, \citenamefont {Yu}, \citenamefont {Zhu}, \citenamefont {Berkelbach}, \citenamefont {Sharma}, \citenamefont {Sokolov},\ and\ \citenamefont {Chan}}]{10.1063/5.0006074}%
  \BibitemOpen
  \bibfield  {author} {\bibinfo {author} {\bibfnamefont {Q.}~\bibnamefont {Sun}}, \bibinfo {author} {\bibfnamefont {X.}~\bibnamefont {Zhang}}, \bibinfo {author} {\bibfnamefont {S.}~\bibnamefont {Banerjee}}, \bibinfo {author} {\bibfnamefont {P.}~\bibnamefont {Bao}}, \bibinfo {author} {\bibfnamefont {M.}~\bibnamefont {Barbry}}, \bibinfo {author} {\bibfnamefont {N.~S.}\ \bibnamefont {Blunt}}, \bibinfo {author} {\bibfnamefont {N.~A.}\ \bibnamefont {Bogdanov}}, \bibinfo {author} {\bibfnamefont {G.~H.}\ \bibnamefont {Booth}}, \bibinfo {author} {\bibfnamefont {J.}~\bibnamefont {Chen}}, \bibinfo {author} {\bibfnamefont {Z.-H.}\ \bibnamefont {Cui}}, \bibinfo {author} {\bibfnamefont {J.~J.}\ \bibnamefont {Eriksen}}, \bibinfo {author} {\bibfnamefont {Y.}~\bibnamefont {Gao}}, \bibinfo {author} {\bibfnamefont {S.}~\bibnamefont {Guo}}, \bibinfo {author} {\bibfnamefont {J.}~\bibnamefont {Hermann}}, \bibinfo {author} {\bibfnamefont {M.~R.}\ \bibnamefont {Hermes}}, \bibinfo {author} {\bibfnamefont {K.}~\bibnamefont {Koh}},
  \bibinfo {author} {\bibfnamefont {P.}~\bibnamefont {Koval}}, \bibinfo {author} {\bibfnamefont {S.}~\bibnamefont {Lehtola}}, \bibinfo {author} {\bibfnamefont {Z.}~\bibnamefont {Li}}, \bibinfo {author} {\bibfnamefont {J.}~\bibnamefont {Liu}}, \bibinfo {author} {\bibfnamefont {N.}~\bibnamefont {Mardirossian}}, \bibinfo {author} {\bibfnamefont {J.~D.}\ \bibnamefont {McClain}}, \bibinfo {author} {\bibfnamefont {M.}~\bibnamefont {Motta}}, \bibinfo {author} {\bibfnamefont {B.}~\bibnamefont {Mussard}}, \bibinfo {author} {\bibfnamefont {H.~Q.}\ \bibnamefont {Pham}}, \bibinfo {author} {\bibfnamefont {A.}~\bibnamefont {Pulkin}}, \bibinfo {author} {\bibfnamefont {W.}~\bibnamefont {Purwanto}}, \bibinfo {author} {\bibfnamefont {P.~J.}\ \bibnamefont {Robinson}}, \bibinfo {author} {\bibfnamefont {E.}~\bibnamefont {Ronca}}, \bibinfo {author} {\bibfnamefont {E.~R.}\ \bibnamefont {Sayfutyarova}}, \bibinfo {author} {\bibfnamefont {M.}~\bibnamefont {Scheurer}}, \bibinfo {author} {\bibfnamefont {H.~F.}\ \bibnamefont {Schurkus}},
  \bibinfo {author} {\bibfnamefont {J.~E.~T.}\ \bibnamefont {Smith}}, \bibinfo {author} {\bibfnamefont {C.}~\bibnamefont {Sun}}, \bibinfo {author} {\bibfnamefont {S.-N.}\ \bibnamefont {Sun}}, \bibinfo {author} {\bibfnamefont {S.}~\bibnamefont {Upadhyay}}, \bibinfo {author} {\bibfnamefont {L.~K.}\ \bibnamefont {Wagner}}, \bibinfo {author} {\bibfnamefont {X.}~\bibnamefont {Wang}}, \bibinfo {author} {\bibfnamefont {A.}~\bibnamefont {White}}, \bibinfo {author} {\bibfnamefont {J.~D.}\ \bibnamefont {Whitfield}}, \bibinfo {author} {\bibfnamefont {M.~J.}\ \bibnamefont {Williamson}}, \bibinfo {author} {\bibfnamefont {S.}~\bibnamefont {Wouters}}, \bibinfo {author} {\bibfnamefont {J.}~\bibnamefont {Yang}}, \bibinfo {author} {\bibfnamefont {J.~M.}\ \bibnamefont {Yu}}, \bibinfo {author} {\bibfnamefont {T.}~\bibnamefont {Zhu}}, \bibinfo {author} {\bibfnamefont {T.~C.}\ \bibnamefont {Berkelbach}}, \bibinfo {author} {\bibfnamefont {S.}~\bibnamefont {Sharma}}, \bibinfo {author} {\bibfnamefont {A.~Y.}\ \bibnamefont
  {Sokolov}},\ and\ \bibinfo {author} {\bibfnamefont {G.~K.-L.}\ \bibnamefont {Chan}},\ }\bibfield  {title} {\bibinfo {title} {{Recent developments in the PySCF program package}},\ }\href {https://doi.org/10.1063/5.0006074} {\bibfield  {journal} {\bibinfo  {journal} {The Journal of Chemical Physics}\ }\textbf {\bibinfo {volume} {153}},\ \bibinfo {pages} {024109} (\bibinfo {year} {2020})},\ \Eprint {https://arxiv.org/abs/https://pubs.aip.org/aip/jcp/article-pdf/doi/10.1063/5.0006074/16722275/024109\_1\_online.pdf} {https://pubs.aip.org/aip/jcp/article-pdf/doi/10.1063/5.0006074/16722275/024109\_1\_online.pdf} \BibitemShut {NoStop}%
\bibitem [{\citenamefont {Hess~Jr}\ \emph {et~al.}(1986)\citenamefont {Hess~Jr}, \citenamefont {Schaad}, \citenamefont {Carsky},\ and\ \citenamefont {Zahradnik}}]{hess1986ab}%
  \BibitemOpen
  \bibfield  {author} {\bibinfo {author} {\bibfnamefont {B.~A.}\ \bibnamefont {Hess~Jr}}, \bibinfo {author} {\bibfnamefont {L.~J.}\ \bibnamefont {Schaad}}, \bibinfo {author} {\bibfnamefont {P.}~\bibnamefont {Carsky}},\ and\ \bibinfo {author} {\bibfnamefont {R.}~\bibnamefont {Zahradnik}},\ }\bibfield  {title} {\bibinfo {title} {Ab initio calculations of vibrational spectra and their use in the identification of unusual molecules},\ }\href@noop {} {\bibfield  {journal} {\bibinfo  {journal} {Chemical Reviews}\ }\textbf {\bibinfo {volume} {86}},\ \bibinfo {pages} {709} (\bibinfo {year} {1986})}\BibitemShut {NoStop}%
\bibitem [{\citenamefont {Maslen}\ \emph {et~al.}(1992)\citenamefont {Maslen}, \citenamefont {Handy}, \citenamefont {Amos},\ and\ \citenamefont {Jayatilaka}}]{maslen1992higher}%
  \BibitemOpen
  \bibfield  {author} {\bibinfo {author} {\bibfnamefont {P.~E.}\ \bibnamefont {Maslen}}, \bibinfo {author} {\bibfnamefont {N.~C.}\ \bibnamefont {Handy}}, \bibinfo {author} {\bibfnamefont {R.~D.}\ \bibnamefont {Amos}},\ and\ \bibinfo {author} {\bibfnamefont {D.}~\bibnamefont {Jayatilaka}},\ }\bibfield  {title} {\bibinfo {title} {Higher analytic derivatives. iv. anharmonic effects in the benzene spectrum},\ }\href@noop {} {\bibfield  {journal} {\bibinfo  {journal} {The Journal of chemical physics}\ }\textbf {\bibinfo {volume} {97}},\ \bibinfo {pages} {4233} (\bibinfo {year} {1992})}\BibitemShut {NoStop}%
\end{thebibliography}%
%
\clearpage
\pagebreak
\setcounter{section}{0}
\setcounter{equation}{0}
\setcounter{figure}{0}
\setcounter{table}{0}
\setcounter{page}{1}
\makeatletter
\renewcommand{\theequation}{S\arabic{equation}}
\renewcommand{\thesection}{S\arabic{section}}
\renewcommand{\thefigure}{S\arabic{figure}}

\title{Supplementary Information: Multi-task learning for molecular electronic structure approaching coupled-cluster accuracy}

\author{Hao Tang}%
 \affiliation{Department of Materials Science and Engineering, Massachusetts Institute of Technology, MA 02139, USA}

\author{Brian Xiao}%
 \affiliation{Department of Physics, Massachusetts Institute of Technology, MA 02139, USA}

\author{Wenhao He}%
   \affiliation{
   The Center for Computational Science and Engineering, Massachusetts Institute of Technology, Cambridge, MA 02139, USA}

\author{Pero Subasic}%
\affiliation{
   Honda Research Institute USA, Inc., 
   San Jose, CA 95134, USA}

\author{Avetik R. Harutyunyan}%
\affiliation{
   Honda Research Institute USA, Inc., 
   San Jose, CA 95134, USA}

\author{Yao Wang}%
\affiliation{
   Department of Chemistry, Emory University, Atlanta, GA 30322, USA}

\author{Fang Liu}%
\affiliation{
   Department of Chemistry, Emory University, Atlanta, GA 30322, USA}

\author{Haowei Xu}%
\email{haoweixu@mit.edu}
\affiliation{
   Department of Nuclear Science and Engineering, Massachusetts Institute of Technology, Cambridge, MA 02139, USA}

\author{Ju Li}%
 \email{liju@mit.edu}
 \affiliation{Department of Materials Science and Engineering, Massachusetts Institute of Technology, MA 02139, USA}
\affiliation{
   Department of Nuclear Science and Engineering, Massachusetts Institute of Technology, Cambridge, MA 02139, USA}

\maketitle

\section{Perturbation theory-based back-propagation}
In the EGNN training, gradient of the loss function to the model parameters needs to be calculated. Gradient back-propagation schemes are well-developed for all computation steps except solving the Schrodinger equation Eq.~3 in the main text. The gradients are numerically unstable when there are near-degenerate energy levels, which is usually the case in molecules. Here, we first use quantum perturbation theory to obtain the first-order change of energy levels and molecular orbitals:
\begin{equation}
\begin{aligned}
    \delta \epsilon_i &= (\mathbf c^i)^\dagger  \delta H^{\rm eff}  \mathbf c^i\\
    \delta \mathbf c^i &= \sum_{p\neq i} \frac{(\mathbf c^p)^\dagger \delta H^{\rm eff} \mathbf c^i}{\epsilon_i - \epsilon_p}\mathbf c^p
\end{aligned}
\end{equation}
Then, we have the gradients to model parameters as:
\begin{equation}
\begin{aligned}
    \nabla_\theta \epsilon_i  &= (\mathbf c^i)^\dagger  (\nabla_\theta \mathbf{V}^\theta ) \mathbf c^i\\
    \nabla_\theta \mathbf c^i &= \sum_{p\neq i} \frac{(\mathbf c^p)^\dagger (\nabla_\theta \mathbf{V}^\theta ) \mathbf c^i}{\epsilon_i - \epsilon_p}\mathbf c^p
    \label{eq:grad}
\end{aligned}
\end{equation}
Using these equations, we derive the gradients of each molecule properties in Eq.~4 as follow:
\begin{equation}
    \begin{aligned}
        \nabla_\theta f_E &= 2\sum_{i=1}^{n_e/2}\nabla V_{ii} \\
        \nabla_\theta f_{\vec{p}} &= -4e\sum_{i=1}^{n_e/2}\sum_{a=n_e/2+1}^{N_{\rm basis}} \text{Re}\frac{\nabla V_{ai}}{\epsilon_i - \epsilon_a}
        \langle\psi_i|\hat{\vec{r}}|\psi_a\rangle  \\
        \nabla_\theta  f_{\mathbf Q} &= 4e^2\sum_{i=1}^{n_e/2}\sum_{a=n_e/2+1}^{N_{\rm basis}} \text{Re}\frac{\nabla V_{ai}}{\epsilon_i - \epsilon_a}\langle\psi_i|\hat{\vec{r}}\hat{\vec{r}}|\psi_a\rangle\\
        \nabla_\theta f_{C_I} &= -4e\sum_{i=1}^{n_e/2}\sum_{a=n_e/2+1}^{N_{\rm basis}} \text{Re}\frac{\nabla V_{ai} (\tilde{\mathbf c}^i)^\dagger (I_I S)\tilde{\mathbf c}^a}{\epsilon_i - \epsilon_a}\\
        \nabla_\theta f_{B_{IJ}} &= 4\sum_{i=1}^{n_e/2}\sum_{a=n_e/2+1}^{N_{\rm basis}} \text{Re}\frac{\nabla V_{ai}}{\epsilon_i - \epsilon_a}\\
        &\times (\tilde{\mathbf c}^i)^\dagger (SI_JPSI_I+SI_IPSI_J)\tilde{\mathbf c}^a
    \end{aligned}
    \label{eq:grad_ground}
\end{equation}
where $\nabla V_{ai}\equiv (\tilde{\mathbf c}^a)^\dagger (\nabla_\theta \mathbf{V}^\theta) \tilde{\mathbf c}^i$, the $N_{\rm basis}\times N_{\rm basis}$ matrix $I_J$ is identity in the block diagonal part of atom $J$ and zero elsewhere. Meanwhile, we define $P\equiv 2\sum_{i=1}^{n_e/2}\tilde{\mathbf c}^i (\tilde{\mathbf c}^i)^\dagger$. The essential method to avoid numerical instability is to remove terms that can be proved to cancel each other. Taking $\nabla_\theta f_{\vec{p}}$ as an example: in Eq.~\eqref{eq:grad}, the summation over $m$ goes through all states except $n$. But as the summed formula is antisymmetric to $m$ and $n$, the terms that $m$ goes from 1 to $n_e/2$ cancel each other. Only terms that $m$ goes from $n_e/2+1$ to $N_{\rm basis}$ have a non-zero contribution to the final gradient. Therefore, $n$ is always occupied, and $m$ is always unoccupied in the summation. As close-shell molecules always have a finite bandgap, $\epsilon_n$ and $\epsilon_m$ are not close to each other in any term of the summation, so evaluating Eq.~\ref{eq:grad_ground} is numerically stable. 

\begin{figure*}[t]
\centering
\includegraphics[width=\linewidth]{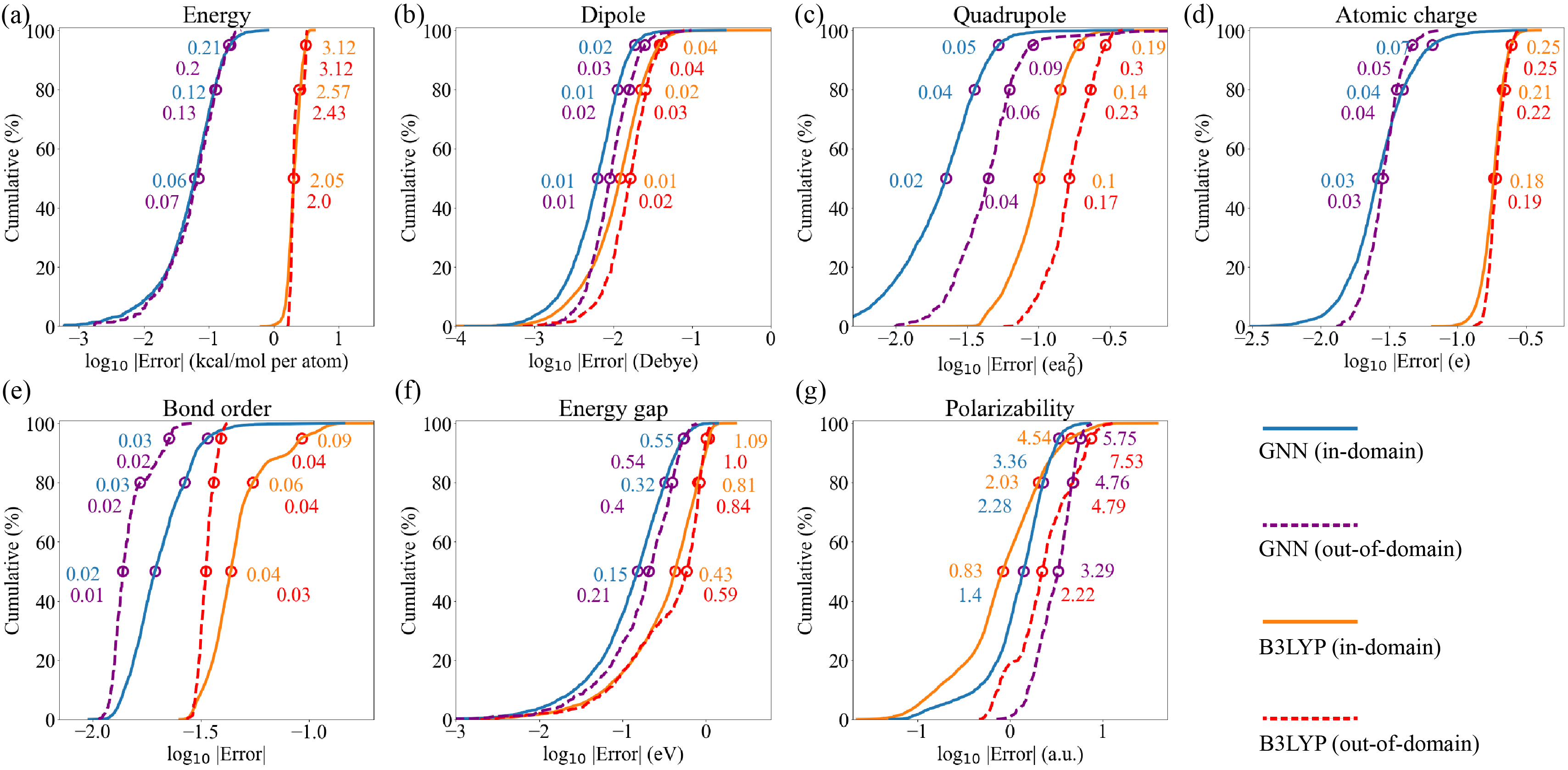}
\caption{The distribution of model prediction accuracy on the test dataset compared to the B3LYP DFT calculations using the CCSD(T) results as the ground truth. a-g Cumulative distribution of prediction errors for the (a) energy, (b) electric dipole moment, (c) electric quadrupole moment, (d) Mulliken atomic charge, (e) Mayer bond order, (f) energy gap (1$^{\rm st}$ excitation energy), and (g) static electric polarizability (a.u. means atomic unit). The blue and orange solid lines represent EGNN and B3LYP results on the in-domain testing dataset, and the purple and red dashed lines represent GNN and B3LYP results on the out-of-domain testing dataset, respectively. We denote the model errors at 50\%, 80\%, and 95\% percentile from the bottom to the top by hollow circles. }
\label{fig:benchmark}
\end{figure*}

Similarly, the gradients of $E_{\rm g}$ and $\alpha$ are as follow:
\begin{equation}
    \begin{aligned}
        \nabla_\theta f_{E_{\rm g}} &= (1+G_1)\left[  \nabla V_{n_e/2+1,n_e/2+1}-\nabla V_{n_e/2,n_e/2}\right]\\
        &+ (\epsilon_{n_e/2+1}-\epsilon_{n_e/2})\nabla_\theta G_1 + \nabla_\theta G_2
    \end{aligned}
\end{equation}
To calculate the gradient of $\alpha$, we first evaluate the gradient of $\alpha_0$, and then derive $\nabla_\theta f_\alpha$ using the chain rule:
\begin{equation}
    \begin{aligned}
        \nabla_\theta \alpha_0 &= 2e^2\sum_{a=n_e/2+1}^{N_{\rm basis}}\sum_{i = 1}^{n_e/2}\text{Re}\left\{\frac{\vec{r}_{ai}\vec{r}_{ia}(\nabla V_{ii}-\nabla V_{aa})}{(\epsilon_a - \epsilon_i)^2}\right.\\
        &\left.-2\sum_{p\neq a,i} \frac{\vec{r}_{ai}}{(\epsilon_a - \epsilon_i)}\left[\frac{\vec{r}_{ip}\nabla V_{pa}}{(\epsilon_p - \epsilon_a)} + \frac{\vec{r}_{pa}\nabla V_{ai}}{(\epsilon_p - \epsilon_i)}\right]\right\}\\
        \nabla_\theta f_\alpha &= (I+\alpha_0 T)^{-1}(\nabla_\theta\alpha_0) (I-T\alpha)\\
        &-\alpha_0(\nabla_\theta T)(I+\alpha_0T)^{-1}\alpha
    \end{aligned}
\end{equation}
The above equations give gradients of all terms in the loss function expressed by gradients to the direct outputs of the EGNN, $\nabla V$, $\nabla_\theta \mathbf G$, and $\nabla_\theta T$.

\begin{table*}
\renewcommand\arraystretch{1.4}
\caption{List of molecule names and number of atomic configurations (labelled in the superscript) for each molecule in the training and testing dataset.}
\begin{tabular}{ p{2.6cm}p{2.8cm}p{3.2cm}p{2.8cm}p{2.8cm}p{2.8cm}}
 \hline
 \hline
Chemical formula & Molecule 1 & Molecule 2 & Molecule 3 & Molecule 4 & Molecule 5\\
 \hline
CH$_4$  & Methane$^{500}$&  --& --& --& --\\
\hline
C$_2$H$_2$   & Acetylene$^{500}$ & --& --& --&   --\\
\hline
C$_2$H$_4$   &  Ethylene$^{500}$ & --& --& --&  --\\
\hline
C$_2$H$_6$   &  Ethane$^{500}$ &  --& --& --&  --\\
\hline
C$_3$H$_4$     &  Propyne$^{300}$ & Allene$^{100}$ & Cyclopropene$^{100}$ & -- & -- \\
 \hline
C$_3$H$_6$     &  Propylene$^{260}$ & Cyclopropane$^{250}$ & --& --&  --\\
 \hline
C$_3$H$_8$    & Propane$^{500}$& --& --& --&  --\\
 \hline
C$_4$H$_6$    & 1,2-Butadiene$^{100}$ & 1,3-Butadiene$^{100}$& 1-Butyne$^{100}$& 2-Butyne$^{100}$&  1-Methylcyclopropene$^{100}$\\
 \hline
C$_4$H$_8$    & Isobutylene$^{100}$ & Cyclobutane$^{100}$& 1-Butene$^{100}$& 2-Butene$^{100}$&  Methylcyclopropane$^{100}$\\
 \hline
C$_4$H$_{10}$    & Butane$^{250}$ & Isobutane$^{250}$& -- & -- & -- \\
 \hline
C$_5$H$_8$    & Isoprene$^{100}$ & Cyclopentene$^{100}$& 1-Pentyne$^{100}$& Methylene-cyclobutane$^{100}$&  1,3-Pentadiene$^{100}$\\
 \hline
C$_5$H$_{10}$    & Cyclopentane$^{100}$ & 1-Pentene$^{100}$& 2-Methyl-1-Butene$^{100}$& 2-Methyl-2-Butene$^{100}$&  3-Methyl-1-Butene$^{100}$\\
 \hline
C$_5$H$_{12}$    &  Neopentane$^{200}$ & Isopentane$^{200}$ & Pentane$^{100}$ & -- & --  \\
 \hline
C$_6$H$_6$  & Benzene$^{100}$ & 1,5-Hexadiyne$^{100}$& 2,4-Hexadiyne$^{100}$& Divinylacetylene$^{100}$&  3,4-Dimethylene-cyclobut-1-ene$^{100}$ \\
 \hline
C$_6$H$_8$  & 1,3-Cyclohexadiene$^{100}$ & 1,4-Cyclohexadiene$^{100}$& Hexa-1,3,5-triene$^{100}$& Methyl-cyclopentadiene$^{100}$&  Divinylethylene$^{100}$\\
 \hline
C$_6$H$_{12}$ & Methyl-cyclopentane$^{100}$ & Cyclohexane$^{100}$& 1-Hexene$^{100}$& cis-4-Methyl-2-pentene$^{100}$&  2-Methyl-1-Pentene$^{100}$\\
 \hline
C$_6$H$_{14}$  & 2,2-Dimethylbutane$^{100}$ & 2,3-Dimethylbutane$^{100}$& 3-Methylpentane$^{100}$& 2-Methylpentane$^{100}$&  Hexane$^{100}$ \\
 \hline
C$_7$H$_8$    & Toluene$^{100}$ & 2,5-Norbornadiene$^{100}$& Quadricyclane$^{100}$& 1,6-Heptadiyne$^{100}$&  Cycloheptatriene$^{100}$  \\
 \hline
C$_7$H$_{10}$ & Norbornene$^{100}$ & 1,3-Cycloheptadiene$^{100}$& 1-Methyl-1,3-cyclohexadiene$^{100}$ & 2-Methyl-1,3-cyclohexadiene$^{100}$&  3-Methylenecyclohexene$^{100}$ \\
 \hline
C$_7$H$_{14}$ & Methyl-cyclohexane$^{50}$ & Cycloheptane$^{50}$& 1-Heptene$^{50}$ & (E)-4,4-Dimethyl-2-pentene$^{50}$&  trans-3-Heptene$^{50}$\\
 \hline
C$_8$H$_8$  & Styrene$^{100}$ & Benzocyclobutene$^{100}$& Cubane$^{100}$ & Semibullvalene$^{100}$&  Cyclooctatetraene$^{100}$\\
 \hline
C$_8$H$_{14}$ & Bimethallyl$^{25}$ & Diisocrotyl$^{50}$& 1,7-Octadiene$^{25}$ & CYCLOOCTENE$^{25}$&  (4E)-2,3-dimethylhexa-1,4-diene$^{25}$\\
 \hline
C$_{10}$H$_{10}$ & 1,3-Divinylbenzene$^{20}$& Diolin$^{20}$& 1,4-Divinylbenzene$^{20}$ & Divinylbenzene$^{20}$&  4-Phenyl-1-butyne$^{20}$\\

 \hline
 \hline
\end{tabular}
\label{table:molecules}
\end{table*}

\begin{table*}
\renewcommand\arraystretch{1.4}
\caption{List of the names and serial numbers (superscript, defined in Ref.~\cite{slayden2001energetics}) of aromatic molecules in the main text Fig.~3.}
\begin{tabular}{ p{2.4cm}p{3.7cm}p{3.4cm}p{4cm}p{3.7cm}}
 \hline
 \hline
Molecule index & 1 & 2 & 3 & 4 \\
 \hline
Name$^{\rm serial\ number}$  & trans-10b,10c-dimethyl-10b,10c-dihydropyrene$^{12}$&  anthracene$^{20}$ & benzo[c]phenanthrene$^{24}$ & 5-ring phenacene, picene$^{28}$ \\
\hline
Molecule index   & 5 & 6 & 7 & 8\\
\hline
Name$^{\rm serial\ number}$   &  Pyrene$^{32}$ & Coronene$^{36}$ & 1,4:2,5-[2.2.2.2]cyclophane$^{40}$ & 9,9'-bianthryl$^{44}$ \\
\hline
Molecule index   & 9 & 10 & 11 & --\\
\hline
Name$^{\rm serial\ number}$   &  $p$-terphenyls$^{48}$ & acenaphthene$^{64}$ & Aceplaidylene$^{68}$ & \\
 \hline
 \hline
\end{tabular}
\label{table:aromatic}
\end{table*}

\section{Dataset and training parameters}

The training domain and out-of-distribution testing testing dataset include 20 and 3 different chemical formula shown as the horizontal axis labels of the first 20 and last 3 columns in Fig.~1c in the main text, respectively. Each chemical formula includes up to 5 different molecules (conformers) taken from the PubChem database. The total number of molecules in the training domain and out-of-distribution testing dataset is 70 and 15, respectively.   The full list of molecules and the number of atomic configurations for each molecule are listed in Table~\ref{table:molecules} in this file.

Each molecule structure is then set as the initial configuration of a MD simulation. The MD simulation uses PreFerred Potential v4.0.0~\cite{takamoto2022towards} at a temperature of 2000 K that enables large bond distortion but does not break the bonds. Initial velocity is set as Maxwell Boltzmann distribution with the same temperature of 2000 K. Langevin NVT dynamics is used with the friction factor of 0.001 fs$^{-1}$ and timestep of 2 fs, and one atomic configuration is sampled every 200 timesteps in the MD trajectory. 500 configurations (including the inital equilibrium configuration) are sampled for each chemical formula in the training domain, where 3/4 of the 10,000 configurations are sampled to form the training dataset, and the left 1/4 forms the in-domain testing dataset. The out-of-distribution testing dataset contains 500 configurations. 

A CCSD(T) calculation with cc-pVTZ basis set is implemented in ORCA~\cite{neese2020orca} for each selected configuration, giving the training labels of total energy, electric dipole and quadrupole moment, Mulliken atomic charge, and Mayer bond order. An EOM-CCSD calculation with cc-pVDZ basis set is then implemented to obtain the first excitation energy (bandgap). Finally, we conduct a polarizability calculation with CCSD and cc-pVDZ basis set. The overlap matrix $S$ and Effective Hamiltonian $F$ is obtained from a BP86 DFT calculations with cc-pVDZ basis set. 

For comparison, B3LYP hybrid DFT calculations are implemented with def2-SVP basis set in ORCA. DSD-PBEP86 double-hybrid DFT calculations are implemented with the def2-TZVP basis set in ORCA. The DM21 double-hybrid DFT calculations are implemented with the def2-TZVP basis set in PySCF program package~\cite{10.1063/5.0006074} The AIQM1 calculations are implemented in the MLatom platform~\cite{dral2024mlatom}. For DM21 and AIQM1, the isolated-atom energies of carbon and hydrogen are re-calibrated according to the least-mean-square criterion to give optimal combination energies in our dataset.

The weight parameters in the loss function is listed as follow: 
$w_V = 0.1$, $w_E=1$, $w_{\vec{p}} = 0.2$, $w_{\mathbf Q}=0.01$, $w_C=0.01$, $w_B=0.02$, $w_{E_{\rm g}}=0.2$, $w_\alpha=3\times 10^{-5}$. All quantities are in atomic unit. The model training is implemented by full gradient descend (FGD) with Adam optimizer. For the finally deployed model (7,440 training data points), it is first trained on 1240 data points sampled from the whole dataset for 5000 FGD steps with initial learning rate of 0.01. The learning rate is decayed by a constant factor $\gamma_1 = 10^{-1/10}$ per 500 steps. Then, the model is trained on the whole dataset with 7,440 data points for 6,000 FGD steps with a initial learning rate of $0.001$. For other models trained on smaller dataset in Fig.~2a in the main text, the model is trained for 3,000 FGD steps with initial learning rate of 0.01 decayed by $\gamma_2 = 10^{-1/6}$ per 500 steps. As the model trained on 640 data points do not converge in the 3000-step training, we implement 10,000-step training, initial learning rate of 0.01 decays by $\gamma_1$ every 500 steps in the first 5,000 steps and keeps constant at the last 5,000 steps.

Aromatic molecules in the main text Fig.~3 include all molecules with serial numbers dividable by 4 and enthalpy of formation provided in Ref.~\cite{slayden2001energetics}. Details of these aromatic molecules are listed in Table~\ref{table:aromatic}.

\section{Infrared spectrum}
In order to evaluate the infrared spectrum, we first implement a B3LYP hybrid DFT calculation with def2-TZVP basis set to obtain the vibrational modes and frequency of a benzene molecule. Then, we generate atomic configurations displaced from the equilibrium configuration along each vibrational mode by a displacement of -0.1, -0.05, 0.05, and 0.1 \AA . Our EGNN model is used to evaluate the electric dipole moment at each configuration, and the dipole-moment derivative with respect to each normal coordinate is evaluated by linear regression. The infrared band intensity of fundamental bands are then evaluated following the method in Ref.~\cite{hess1986ab}.

As the two combination bands at 1800 - 2000 cm$^{-1}$ are mainly contributed by $\nu_{10}+\nu_{17}$ and $\nu_{5}+\nu_{17}$~\cite{maslen1992higher}, we generate atomic configurations displaced from the equilibrium configuration by displacement vectors of $0.1(\vec{e}_i+\vec{e}_j), 0.1(\vec{e}_i-\vec{e}_j), 0.1(-\vec{e}_i+\vec{e}_j)$, and $0.1(-\vec{e}_i-\vec{e}_j)$ \AA , where $(\vec{e}_i, \vec{e}_j)$ are the pair of vibrational modes contribution to each combination band. The second-order dipole-moment derivatives with respect to each pair of normal coordinates $\frac{\partial^2\vec{p}}{\partial Q_i\partial Q_j}$ are then obtained by finite difference method. The leading-order anharmonic constants are also evaluated by finite difference method. These parameters are then used to calculate intensity of the combination bands by Fermi's golden rule. Using the calculated infrared spectrum peak positions and intensity, we add Gaussian broadening to each peak and fit their bandwidth to the experimental spectrum.

\end{document}